\documentclass[prd,nofootinbib,floats,superscriptaddress,eqsecnum,tightenlines,preprintnumbers,11pt]{revtex4}

\usepackage{mathrsfs}
\usepackage{xspace}
\usepackage{mathtools, amssymb}
\usepackage{amsmath,scalerel}
\usepackage{hyperref}
\usepackage{graphicx}
\usepackage{amsmath,amssymb,amsfonts,amsthm,latexsym}
\usepackage{marginnote}
\usepackage{xcolor}
\usepackage{physics}
\usepackage{nicefrac}
\usepackage{stmaryrd}
\usepackage{amsmath}
\usepackage{amsfonts}
\usepackage{amssymb}
\usepackage{graphicx}
\usepackage{bbm}
\usepackage{mathrsfs}
\usepackage{xspace}
\usepackage{mathtools, amssymb}
\let\curv\mathscr
\usepackage{tikz}
\tikzset{every picture/.style={line width=0.75pt}}

\begin{document}

\title{Rational regular black holes in non-polynomial gravity}

\author{Aimeric Coll\'{e}aux}
\affiliation{Institute of Theoretical Physics, Faculty of Mathematics and  Physics, Charles University,  V Hole\v{s}ovi\v{c}k\'{a}ch  2,  Prague  180  00,  Czech  Republic}
\email{aimeric.colleaux@matfyz.cuni.cz}
\date{\today}

\begin{abstract}
We consider the $d\geq 4$ non-polynomial pure gravity theories defined so that their 2D reduction to spherical symmetry minimally modifies that of General Relativity at small distances. They satisfy Birkhoff theorem and admit exact black hole solutions for an arbitrary number of couplings. These solutions have a logarithmic leading correction to the entropy and a quantum-like correction to the Newtonian potential, both arising from the regularised critical order $d=2p$ theory, which is related in even dimensions to the Euler characteristic of the 2D orbit space. These black holes can be made regular, without the need for infinite towers of corrections, using invariants of any order $p>d/2$. In even dimensions, their smoothness is linear in the number of corrections and does not require fine-tuning. We construct a class of theories admitting as unique (massless) vacuum an extremal black hole deformation of (A)dS$_4$ or M$_4$ with vanishing entropy, so that Nernst's third law of thermodynamics is satisfied without discontinuity in the entropy.  We show that a Maxwell field does not disrupt the regularity of the metric, and that simple non-minimal couplings in the action enable to regularise the electric field. Considering infinite towers of corrections, we construct quasi-regular black holes with one horizon as well as regular (A)dS-core ones with a near-extremal inner horizon for any mass. When charged, these two types of black holes prevent or tame the mass inflation provided the mass and charge satisfy an inequality of the form $M > \alpha \ell + \beta Q^2 / \ell$, where $\alpha$ and $\beta$ are numerical factors depending on the specific model, while $\ell$ is the length scale controlling the high-energy corrections. The quasi-regular black holes are sufficient to resolve the Coulomb singularity without the need for non-minimal couplings.
\end{abstract}

\maketitle


\section{Introduction}

Although General Relativity (GR) continues to enjoy great experimental success \cite{LIGOScientific:2016aoc,LIGOScientific:2017vwq,EventHorizonTelescope:2019dse,EventHorizonTelescope:2022wkp}, it also suffers from internal inconsistencies, such as those related to the presence of singularities in its solutions under very generic physical conditions \cite{Penrose:1964wq,Hawking:1970zqf}, in the form of geodesic incompleteness, but also curvature divergences, infinite tidal forces and topological defects \cite{Ellis:1977pj,Joshi:2013xoa,LimaJunior:2025uyj,Magalhaes:2024smm}. This lack of predictivity at short scales is expected to be resolved by a quantum theory of gravity, but no consensus has yet been reached among the existing proposals \cite{Reuter:1996cp,Eichhorn:2017egq,Eichhorn:2026uqj,Ambjorn:1991pq,Rovelli:1994ge,Perez:2012wv,Gross:1986iv,Blau:2009fzi,Modesto:2011kw,Anselmi:2017ygm}.  Furthermore, the mechanisms to avoid singularities being very different in each of these, effective approaches, more agnostic on the specific UV completion, are important to construct and study. Many models admitting regular black holes by modifying the matter content have been proposed, involving for instance gauge, scalar and vector fields, modified gravities with matter, as well as effective anisotropic fluids \cite{Ayon-Beato:1998hmi, Balakin:2015gpq, Balakin:2016mnn, Nojiri:2017kex, Chinaglia:2018uol,Bronnikov:2005gm, Chamseddine:2016ktu, BenAchour:2017ivq, Bakopoulos:2023fmv,Colleaux:2020wfv, Fernandes:2025fnz, Charmousis:2025jpx, Fernandes:2025mic,Eichhorn:2025pgy,Junior:2026ism, Olmo:2012nx, Bambi:2015zch,Olmo:2015bya,Olmo:2022cui,Menchon:2017qed,Lessa:2024erf}. However, if the resolution of singularities comes from a pure quantum gravity theory and survives in such effective regimes, one could expect instead that geometrical corrections should cure singularities independently from the matter content, and that this property should be preserved when physically relevant matter is included in the system\footnote{One may expect it to be the case for approaches such as Loop Quantum Gravity, Asymptotic Safety and Causal Dynamical Triangulation, which quantize GR in vacuum.}. If so, some diffeomorphism invariant effective field theory of gravity with Lagrangian density $\sum  \mathcal{R}_p\left(g,\partial\right)\ell^{2p}$, for some length scale $[\ell]=L$, typically the Planck length, and curvature invariants $\mathcal{R}_p$ of order $p$, $[\mathcal{R}_p]=L^{-2p}$,  describing strong field corrections controlled by $\ell$, should yield regular black holes and reduce to GR for $\ell=0$. As the only curvature corrections to GR with second order field equations for all metric fields are the dimensionally-continued Euler densities \cite{Lovelock:1971yv}, which do not affect the dynamics in 4D beyond GR \cite{Lovelock:1972vz} and do not resolve the black hole singularity in higher dimensions \cite{Maeda:2011ii}, the effective field equations are expected to be higher-order\footnote{This is more generally true for quantum effective actions, but as the Euler-Heisenberg theory shows, these can still have second order field equations for some fields and under certain approximations.}. Nonetheless, many black holes and cosmologies obtained from some approximations of quantum gravity (inspired) approaches do not involve additional charges and integration constants \cite{Bonanno:2000ep, Singh:2006im, Modesto:2008im,Peltola:2008pa, Peltola:2009jm, Bambi:2013caa, Torres:2017ygl, Wang:2019dcj,Platania:2019kyx,Kelly:2020uwj,Gambini:2020nsf,Basile:2021euh} and so might be viewed as vacuum solutions of some effective theories with second order field equations (at least) in spherical symmetry. This property being also of great help to obtain exact solutions and analytical results makes it a natural approximation for effective theories as well. To the best of our knowledge, the first gravitational theories with these properties admitting regular black holes and non-singular cosmological solutions have been obtained in \cite{Colleaux:2017ibe} and \cite{Colleaux:2019ckh} in four and higher dimensions using polynomials of non-analytic curvature invariants\footnote{Although regular black holes had been reported in conformal gravities \cite{Bambi:2016wdn}, these theories are not EFTs, as noted there.}.

\medskip

In this paper, we summarise and extend the results of (mostly Chap.III) of \cite{Colleaux:2019ckh} regarding these pure gravity theories with regular black hole solutions. They are inspired by cosmological gravity models \cite{Date:2008gq, Helling:2009ia}, quasi-topological (QTG) \cite{Oliva:2010eb,Myers:2010ru,Oliva:2011xu,Hennigar:2017ego, Ahmed:2017jod,Cisterna:2017umf} and non-polynomial gravities (NPG) \cite{Deser:2007za,Gao:2012fd,Colleaux:2015yta, Chinaglia:2017wim}, which admit second order field equations in FLRW cosmologies and spherical symmetry, due to some algebraic properties of the curvature in these geometries \cite{Deser:2005pc}. Thus, they are also naturally related to the effective 2D Einstein-dilaton approach, whose solutions are sometimes interpreted as describing $d$-dimensional spherically symmetric line elements and in particular 4D regular black holes \cite{Ziprick:2010vb,Taves:2014laa,Kunstatter:2015raa,Kunstatter:2015vxa,Carballo-Rubio:2025ntd}. In four dimensions, non-analyticity in the curvature seems necessary to obtain second order field equation in spherical symmetry\footnote{There exists polynomial four-dimensional gravities leading to second order equations for FLRW cosmologies, see \cite{Date:2008gq,Colleaux:2015yta} and more recently \cite{Moreno:2023arp}. However, already static spherical symmetry seems too restrictive to have this property.}. However, we obtained a dictionary between 2D Horndeski models and $d\geq 4$ dimensional non-polynomial gravities \cite{Colleaux:2017ibe,Colleaux:2019ckh}, meaning that the approach is not much constrained as most static black hole metrics can be seen as solutions of some NPG\footnote{For instance, we set up a reconstruction procedure in Chap.IV of \cite{Colleaux:2019ckh}, which, given a static spherically symmetric metric \eqref{Weylgauge}, with $a(r)$ linear in the mass, yields a NPG theory satisfying Birkhoff theorem and admitting that solution. We applied this to Modesto's semi-polymeric black hole \cite{Modesto:2008im}, Visser-Hochberg-Simpson black bounce \cite{Visser:1997yn} and D'Ambrosio-Rovelli metric \cite{DAmbrosio:2018wgv}, what enabled us to construct their 4D charged generalisations and obtain the region of parameter space $\{M,Q\}$ for which they avoid the mass-inflation instability.}. For this reason, it is important to establish a priori criteria, independent from the specific solutions, e.g. akin to energy conditions in the effective fluid approach, which single out some specific 2D Horndeski theories.

Two main classes were constructed in \cite{Colleaux:2017ibe,Colleaux:2019ckh}, assuming that their (2D) Lagrangians and field equations contain the same operators as those appearing in the spherically symmetric decomposition of, respectively, General Relativity and Gauss-Bonnet (or Lovelock) gravity, while constituting genuine high-energy corrections to the former\footnote{Remark that other criteria have been used in \cite{Colleaux:2017ibe,Colleaux:2019ckh} to refine the selection of admissible theories. In particular they should admit both non-singular cosmologies and spherical black holes, as well as regular planar and hyperbolic solutions, and this regularity should be robust to the inclusion of matter fields, such as the EM field. Another selection rule for effective 2D Horndeski and their NPG formulation has recently been proposed in \cite{Thaalba:2026abz},  based on their ability to support dynamical regular matter configurations. As it turns out, the theories we study here satisfy this parity condition.}. This naturally results in quasi-topological gravities, in the sense that Birkhoff theorem is satisfied, single-function solutions exist ($g_{tt} \times g_{rr}=-1$ in Schwarzschild gauge) and the remaining field equation can be integrated once into an algebraic equation, as in GR. From a 2D perspective, they generalise existing GR-like models \cite{Taves:2014laa}, further studied recently in \cite{Barenboim:2025fds, Barenboim:2025ckx} and Lovelock-like ones \cite{Kunstatter:2015raa,Kunstatter:2015vxa}. Interestingly, they both yield regular black holes for very mild fine-tuning of their coupling constants. Indeed, while the resolution of curvature singularities in the former theories is the main topic of the present work, that of the latter follows from \cite{Kunstatter:2015raa,Kunstatter:2015vxa} which have shown that extending the 2D reduction of Lovelock-Lanczos gravity beyond the critical order, $d \leq 2p$, where it usually trivialises, up to infinite order, dubbed Designer Lovelock Gravity (DLG), allows to obtain regular black holes.  They have been used in  Chap.V of \cite{Colleaux:2019ckh} to obtain 4D NPG theories with (generalised) Wheeler polynomial \cite{Wheeler:1985qd} admitting both non-singular black holes and cosmologies, in particular quantum gravity inspired geometries, such as the LQC bounce \cite{Singh:2006im} with its associated black hole \cite{Kelly:2020uwj} and renormalisation-group improved regular black hole \cite{Bonanno:2000ep} with an associated past-eternal quasi de Sitter cosmology.

\medskip 
 
Recent developments in $d\geq 5$ quasi-topological gravities \cite{Bueno:2019ltp,Bueno:2022res,Moreno:2023rfl,Bueno:2025qjk} have enabled to construct regular black holes \cite{Bueno:2024dgm} and cosmologies \cite{Sueto:2026epz} using the above property of infinite Wheeler polynomials, but from analytic functions of the Riemann tensor, which are certainly preferred, allowing in particular the study of gravitational collapse \cite{Bueno:2024zsx, Bueno:2025gjg} and to address the strong coupling issue using infinite derivative gravity \cite{Bueno:2026oyg}. However, the situation remains similar in four dimensions, as there are still no known examples of analytic purely metric EFT  with this property\footnote{While there are analytic 4D theories with spherically symmetry field equations of reduced (third) order, so-called generalised QTG, no regular black hole solutions are known. Furthermore, while (A)dS-core regular solutions have recently been reported in higher-order gravities \cite{Giacchini:2024exc,Giacchini:2025gzw}, there is no indications so far that they describe black holes with Schwarzschild asymptotic behaviour. However, pure gravity regular black holes have been obtained in other context, such as teleparallel gravity \cite{Bohmer:2019vff} and $f(R)$ gravities which are approximatively scale-invariant for small Ricci scalar \cite{Bertipagani:2020awe}.}. Nonetheless, some progress has recently been made, following the renewed interest in NPG and their ability to produce RBH solutions \cite{Frolov:2024hhe,Bueno:2025zaj,Borissova:2026wmn}. For instance, a complete dictionary between 2D Horndeski and $d$-dimensional NPG has been obtained in \cite{Borissova:2026krh}. Furthermore, it was shown in \cite{Bueno:2025zaj} that for (non-conformally flat) perturbations around maximally symmetric spaces, the field equations of a DLG NPG are analytic and second order, like in GR, extending the applicability of NPG beyond spherical symmetry. However, it should be noted that conformally flat solutions can only be obtained as limits due to non-analyticity in the curvature, while Wheeler polynomials precisely have maximally symmetric vacua. By contrast, the vacua of the theories we study here are generically small scale deformations of maximally symmetric spaces and thus are well-defined solutions of the corresponding $d$-dimensional NPG. Other extensions of 4D NPG beyond spherical symmetry have been obtained in \cite{Colleaux:2026qew} and were shown to admit exact stationary solutions with NUT charge, or describing the near-horizon geometries of extremal rotating black holes \cite{Kunduri:2007vf} and swirling universes \cite{Astorino:2022aam,Barrientos:2024umq}, all related by double Wick rotations \cite{Colleaux:2025uiw}. Further developments of NPG and QTG in spherical symmetry include the study of collapse, radiating solutions, wormholes, black bounces, extremal black holes, thermodynamics, cosmology \cite{Frolov:2026tft, Mazza:2026taj,Borissova:2026prd, Borissova:2026rbi,Saridakis:2026oyf}.

\medskip

Our aim here is to investigate the theory modelled after the spherically symmetric reduction of GR \cite{Colleaux:2017ibe} and show that it satisfies some interesting properties, some similar to Lovelock gravity and DLG, e.g. the critical order theory $d=2p$ yields in spherical symmetry the Euler characteristic of the 2D orbit space and the entropy is the same as that of DLG. Interestingly, when dimensionally regularised in a well-defined way, the critical order theory yields a logarithmic corrections to the entropy and quantum-like correction to Newton's potential, which are both robust quantum gravity predictions  \cite{Donoghue:1993eb,Fursaev:1994te, Kaul:2000kf, Carlip:2000nv, Engle:2009vc, Sen:2012kpz}. There are also some genuine peculiarities such as having regular black holes from theories consisting of finite order polynomials of non-analytic invariants, contrary to those NPG which are DLG and thus require to specify an entire analytic function (of non-analytic invariants) to achieve this goal. Furthermore, just like DLG admit regular black holes under very mild fine-tuning on its (infinite number of) coupling constants, while the smoothness of the solutions must be designed \cite{Bueno:2024dgm}, we show that the theory we study admit smooth even (odd) dimensional black holes for infinite (finite) number of corrections $m$ with similar mild fine-tuning, the smoothness of the solutions depending linearly on $m$ in even-dimensions. It follows from this property that the regularity of black hole solutions is robust when a point-like charge sources the electromagnetic field, as we will show. Another interesting feature is that some of these models admit exact non-conformally flat vacua, which thus bypass the issue of non-analyticity mentioned above. These vacua can be in particular smooth extremal massless black holes, what, we show, implies that Nernst third law of thermodynamics, stating that the entropy of a system must go to zero or to a universal constant as its temperature goes to zero \cite{Wald:1997qp}, is satisfied in the corresponding models, essentially because the location of the degenerate horizon is given by $\ell$, a universal constant in this context.

Most of the regular black holes with spherical horizons mentioned so far have (A)dS-cores and thus necessarily possess an inner Cauchy horizon\footnote{Remark that (A)dS-core hyperbolic and toroidal black holes can be designed to be devoid of Cauchy horizon, in which case they do not suffer from mass inflation, see \cite{Calza:2025mrt}.},  what makes them unstable against the mass inflation instability \cite{Poisson:1989vv, Poisson:1989zz, Ori:1991zz}, protecting them from causality violations, as advocated in \cite{Carballo-Rubio:2024dca}, but preventing these black holes to be both stable and static. This instability arises when outgoing and ingoing perturbations (usually modelled as spherically symmetric null shells, although these are not guarantee to exist in a generic pure gravity theory \cite{DiFilippo:2026jpv}) are taken into account, because null rays are exponentially focused at the inner horizon, what end up triggering an exponential growth of the mass, in the region of the Penrose diagram enclosed by the shells and the inner horizon. We proposed in \cite{Colleaux:2019ckh} and will review here two mechanisms to avoid or tame the mass inflation, based on black holes with a quasi-regular singularity which possess a single horizon in the neutral case, or based on regular black holes whose inner horizon in the non-perturbative limit becomes inner-extremal, a mechanism which has recently been studied in details in \cite{Carballo-Rubio:2022aa, Franzin:2022aa, McMaken:2023aa, DiFilippo:2024mwm}. When charged, these solutions preserve this property when the mass and charge satisfy an inequality of the form $M > \alpha \ell + \beta Q^2 / \ell$, where $\alpha$ and $\beta$ are numerical factors depending on the specific model.

\medskip

The paper is organised as follows. In Sec.\ref{SecMincorrgr} we present the $d$-dimensional non-polynomial gravity and its 2D spherical (planar, hyperbolic) symmetric reduction. In Sec. \ref{BHsol}, we prove Birkhoff theorem and obtain the black hole solutions, which are exact for an arbitrary number of couplings. Using the generalised Misner-Sharp mass and the minisuperspace Wald entropy, studied in the Appendix \ref{App}, see \cite{Colleaux:2019ckh}, it is shown that they satisfy the first law of black hole thermodynamics. Their asymptotic behaviour is analysed and we show that quantum-like corrections to Newton's potential are present in generic dimensions and obtain the few conditions for which the black holes are regular. In Sec.\ref{RRBH}, we analyse the smoothness of these (A)dS-core black holes and their smooth vacua. We construct a class of theories admitting a shared (massless) vacuum and show that when it is a massless extremal black holes, Nernst third law is satisfied. The effect of infinite towers of corrections on the smoothness and mass inflation instability is studied in Sec.\eqref{Sec.inf}. Finally we provide an analysis of the charged black holes and their regularity in Appendix \ref{secCharge}.

\section{Minimal curvature corrections to General Relativity}
\label{SecMincorrgr}

In this section, we first present in \ref{Sec.NPG} the $d=n+2$ dimensional pure gravity theory, introduced in \cite{Colleaux:2017ibe} and further studied in \cite{Colleaux:2019ckh}, which is constructed so that its 2D spherically symmetric reduction generalises that of General Relativity in a minimal way \ref{Sec.2DRed}. The reduced field equations and the generalised Misner-Sharp quasi-local mass are obtained. Finally, the critical order theory and its relation to the 2D Euler characteristic of the orbit space are given in \ref{2dchi}.

\subsection{Non-polynomial gravity action}\label{Sec.NPG}

Let us consider the following pure gravity theory, non-analytic in the curvature, given by the action
 \begin{eqnarray}
 I = I_g +I_b  + I_m  \,,
 \end{eqnarray}
 and associated gravitational field equations 
 \begin{eqnarray}
\mathcal{G}_{\mu\nu} = 8 \pi G T_{\mu\nu} \,,\label{FieldEq}
\end{eqnarray}
 where $I_m$ is the matter action while the gravitational bulk part is given by
\begin{eqnarray}
\label{fullaction}
I_g =\frac{1 }{16\pi G }  \int_\mathcal{M} d^{\, n+2}x  \, \sqrt{-g} \sum _{p=-1}^m \ell^{2p} \,  \mathcal{R}^{p} \, \left(\alpha_{p} \,  \mathcal{R}+ \frac{\beta_{p}}{n-2p} \, \mathcal{S}_{p} \right)  
\end{eqnarray}
and consists in high-energy corrections to General Relativity, controlled by a length scale $\ell$ and two series of dimensionless coupling constants $\alpha$ and $\beta$. It is a polynomial in the following non-analytic curvature invariants, 
\begin{eqnarray}
\label{RS}
\begin{split}
\mathcal{S}_{p} &= \frac{1}{1-n^2} \left(  \frac{4 n  p (1+2p)}{1-n^2} \,\nabla_\gamma u_{\alpha\beta} \nabla^\gamma u^{\alpha\beta}   + R + 2 n \left(R^{\alpha\beta} - \nabla^\alpha \nabla^\beta \right) u_{\alpha\beta} \right),
\\
\\
\mathcal{R} &= \frac{1}{(1-n)^2(1+n)} \Big(n R + 2 \left(R^{\alpha\beta} - \nabla^\alpha \nabla^\beta \right) u_{\alpha\beta}\Big),
\end{split}
\end{eqnarray}
which have the special property that they become analytic and second order in spherical symmetry, as shown below, although they involve the non-analytic curvature tensor $u$, given by 
\begin{eqnarray}
\label{utensor}
u^{\mu}{}_{\nu} =  \left(C_{\sigma\rho\delta} C^{\sigma\rho\delta} \right)^{-1}  \left(  C^{\mu\beta\alpha}  -(1+n) C^{\mu\alpha\beta} \right) C_{\nu \alpha \beta } \; ,
\end{eqnarray} 
in terms of the Cotton tensor, which characterises conformal flatness in three dimensions, 
\begin{eqnarray}
C_{\alpha\beta\gamma} = \nabla_\alpha R_{\beta \gamma} -  \nabla_\beta R_{\alpha \gamma} + \frac{1}{2(1+n)} \left( g_{\alpha \gamma} \nabla_\beta R - g_{\beta \gamma} \nabla_\alpha R \right).
\end{eqnarray} 
Note that we have included the cosmological constant and the four dimensional Ricci scalar into the sum, so that setting $\ell=0$ yields General Relativity. This requires to set 
\begin{eqnarray}
\alpha_{-1} =-2  \ell^{2} \, \Lambda \,,\;\;\;\;\; \beta_{-1}=0  \,,\;\;\;\;\; \alpha_{0}=n(n-1)  \,,\;\;\;\;\; \beta_0=n \,, \label{GRLcc}
\end{eqnarray}
what we will assume throughout this paper, because of the following off-shell identity, 
\begin{eqnarray}
R=n(n-1) \mathcal{R} + n \, \mathcal{S}_{0}.
\end{eqnarray}
Counting the number of derivatives contained in the tensor $u$ as being zero, which is coherent with its scaling property and is exactly true in spherical symmetry, as we will see, the action is overall a (rather exotic) EFT of gravity with each term of the sum being of order $2(p+1)$, i.e. involving $2(p+1)$ derivatives of the metric field\footnote{Except in the abstract and introduction, the label $p$ in the sums \eqref{fullaction} indicates the number of corrections, or loops in a quantum gravity inspired setting, not (half) the order as it is usually done in Lovelock gravity.}.

In order for this theory to have a well-defined variational principle in spherical symmetry, we have supplemented the bulk gravitational action with the following boundary term,
\begin{eqnarray}
I_b =   \frac{\epsilon}{8\pi G }  \oint_\mathcal{\partial M} d^{\, n+1}x \,  \sqrt{|\lambda|}   \, \sum _{p=0}^m \ell^{2p} \, \frac{\beta_{p}}{n-2p} \,  \mathcal{R} ^{p} \,  K \, .  
\label{Ib}
\end{eqnarray}
where $\lambda$ and $K$ are respectively the induced metric  and the extrinsic curvature of the boundary $\partial M$  with time or space-like components respectively satisfying $\epsilon=1,-1$, so that the $p=0$ term corresponds to the Gibbons–Hawking–York boundary term.

It is clear from the definitions of the bulk \eqref{fullaction} and boundary action \eqref{Ib} that, in even dimensions, a divergence occurs for the \textit{critical order} invariant $d=2(p+1)$, or $n=2p$, which contains as many derivatives as the dimension of the manifold. In Lovelock-Lanczos gravity, it corresponds to the Euler density of the manifold, which when integrated yields its Euler characteristic. The well-defined way to treat this invariant and its relation with the 2D Euler characteristic of the orbit space in spherical symmetry will be explained in \ref{2dchi}. 

\subsection{Two-dimensional reduction to spherical symmetry}\label{Sec.2DRed}

\subsubsection{Geometrical setting}

Let us consider the reduction of $\mathcal{M}$ to spherically, hyperbolic and planar symmetric spacetimes, respectively characterised by the discrete parameter $k=1,-1,0$, given by the warped product $\mathcal{M}=\Sigma\times\Omega_r$, with metric 
\begin{eqnarray}
\label{DSS}
ds^2 =d\Sigma^{\,2} +  d\Omega^{\,2}_r
\end{eqnarray}
where
\begin{eqnarray}
d\Sigma^{\,2} = \gamma_{ab} dx^a dx^b \,,\;\;\;\;\;\;\;\;\; d\Omega^{\,2}_r = r^2 d\Omega_{n,k}^{\,2} \,,
\end{eqnarray}
where $r$ is a scalar field on the two-dimensional manifold $\Sigma$ with coordinates $x^a$, $a=1,2$, while $\Omega_{n,k}$ is the $n$-dimensional sphere, hyperbolic and Euclidean planes,
\begin{eqnarray*}
d\Omega_{n,k}^{\,2} = \left\{
  \begin{array}{@{}ll@{}}
 \;\;  d\theta_1^{\,2} + \sum\limits_{i=2}^{n} \prod\limits_{j=1}^{i-1} \sin^2\theta_j d\theta_i^{\,2} \; , \;\;\;\;\;\;\;\;\;\;\;\;\;\;\;\;\;\;\;\;\;\;\;\;\;\;\;\;\,\;\;\; \text{for} \;\; k=1 \,,
 \\  \;\;  \sum\limits_{i=1}^{n} d\phi_i^{\,2} \; ,  \;\;\;\;\;\;\;\;\;\;\;\;\;\;\;\;\;\;\;\;\;\;\;\;\;\;\;\;\,\;\;\;\;\;\;\;\;\,\;\;\;\,\;\;\;\;\,\;\;\;\;\;\;\;\,\,\;\;\; \text{for}  \;\;  k=0 \, , 
  \\  \;\;  d\theta_1^{\,2} + \sinh^2\theta_1 \left( d\theta_2^{\,2} + \sum\limits_{i=3}^{n} \prod\limits_{j=2}^{i-1} \sin^2\theta_j d\theta_i^{\,2} \right) \; , \,\,\,\,\, \text{for}  \;\;  k=-1 \, ,
  \end{array}\right.
\end{eqnarray*}
or their compact counterparts \cite{Vanzo:1997gw}. From this decomposition naturally come associated the Ricci scalars of $\Sigma$ and $\Omega_r$,
\begin{eqnarray}
R\left(\gamma\right) \,,\;\;\;\;\; R\left(\Omega\right) =  \frac{n(n-1) k}{r^2}  \,,
\end{eqnarray}
and the metric-compatible covariant derivative associated with $\gamma_{ab}$, which we denote $D_a$. Furthermore, choosing the boundary to respect the symmetry of \eqref{DSS}, we consider its unit normal $n^\alpha=\{n^a(x), 0, ..., 0 \}$, such that $\gamma^{ab} n_a n_b = \epsilon$. Thus, the $d$-dimensional extrinsic curvature $K$ can be written in terms of the $2$-dimensional one $K^{(2)}$ as
\begin{eqnarray}
K  = K^{(2)} +n  \frac{n^a D_a r}{r} \, . 
\end{eqnarray}
The determinents of the metrics also split into $\sqrt{-g}= \sqrt{-\gamma} \sqrt{\sigma} \,  r^n $ and $\sqrt{|\lambda|}= \sqrt{|h|} \sqrt{\sigma} \, r^n $, where $\sigma$ and $h$ are respectively the determinants of the induced metrics on the horizon manifold of radius unity $\Omega_{n,k}$, and  of the induced metric on the boundary of $\Sigma$, noted $\partial\Sigma$. We define the area of $\Omega_{n,k}$ as
\begin{eqnarray}
V_{n,k} = \int d^n x \sqrt{\sigma} \,,
\end{eqnarray}
which is finite providing the horizon manifolds $\Omega_{n,0}$ and $\Omega_{n,-1}$ are compact.

\subsubsection{Dictionary}\label{Dico}
In this class of spacetimes, the non-analytic curvature tensor $u$ \eqref{utensor} decomposes as \cite{Colleaux:2019ckh}
\begin{eqnarray}
u^{\mu}{}_{\nu} = \frac{1}{2n} \left( -  \delta^{\mu}{}_{\nu} + \left(1-n^2 \right) \gamma^{\mu}{}_{\nu} \right), \label{uDecomp}
\end{eqnarray}
where $\gamma^{\mu}{}_{\nu}$ is the $d$-dimensional projector to the manifold $\Sigma$, see Appendix \ref{appci} where the 2D decomposition of the Cotton tensor and its squares are reported. This relation is identically satisfied for $d=3$ and becomes non-trivial for $d> 3$. As claimed above, we see that this tensor does not contain any derivatives in this class of spacetimes. We have shown in \cite{Colleaux:2017ibe} and \cite{Colleaux:2019ckh} that it enables to obtain a dictionary between two-dimensional scalar-tensor theories and $d$-dimensional non-polynomial gravity. In particular, we are interested here by the following relations for spacetimes of the form \eqref{DSS}, 
\begin{eqnarray}
\label{DefNPGScalars1}
\begin{split} 
 R\left(\Omega \right) &=   R -  \left(R^{\alpha\beta}-\nabla^\alpha \nabla^\beta \right) \gamma_{\alpha\beta} \, ,   \\ 
 R\left(\gamma \right) &= \left(R^{\alpha\beta}+\nabla^\alpha \nabla^\beta \right) \gamma_{\alpha\beta} -\frac{n-1}{2} \nabla_\mu \gamma_{\alpha\beta} \nabla^\mu \gamma^{\alpha\beta}  \, , 
\end{split} 
\end{eqnarray}
and 
\begin{eqnarray}
\label{NPGKinetic}
\nabla_\mu \gamma_{\alpha\beta} \nabla^\mu \gamma^{\alpha\beta} =  2n \, \frac{\gamma^{ab} D_a r \, D_b r }{r^2}  \, , \;\; \;\; 
  \nabla^\alpha \nabla^\beta \gamma_{\alpha\beta} = \frac{D^2 r^n}{r^n}  \,,
\end{eqnarray}
where $\gamma_{\mu\nu}$ can be defined in a $d$-dimensional background independent manner through \eqref{uDecomp} and \eqref{utensor}. Notice that the curvature invariant $\mathcal{R}$ \eqref{RS} which generates the higher-order corrections in the action \eqref{fullaction} is proportional in \eqref{DSS} to the Ricci scalar of the horizon manifold,
\begin{eqnarray}
\mathcal{R}= \frac{R\left(\Omega \right)}{n(n-1)} = \frac{k}{r^2}\,.
\end{eqnarray}
It is important to emphasise that there are infinitely many dictionaries achieving the same goal of lifting 2D curvature invariants to higher dimensions. The only way to distinguish between these is to investigate more general classes of spacetimes than \eqref{DSS} and construct  $d$-dimensional representatives which are analytic and second order for these classes as well, see  the discussion in Chap.II.B.2 of \cite{Colleaux:2019ckh} and the alternatively dictionary recently introduced in \cite{Borissova:2026krh}.

\subsubsection{Definition of the theory}\label{Deftheory}

We have now introduced all the tools required to understand precisely how the action \eqref{fullaction} is constructed. Our aim is to impose some reasonable, and as minimal as possible, requirements on the 2D action to reduce the large freedom of 2D Horndeski gravity, and then to investigate the physical consequences of these choices in the rest of this work. This should be contrasted with the opposite approach of designing black hole metrics and reconstructing the corresponding theories admitting these as solutions. From the existence of dictionaries between 2D Horndeski and NPG \cite{Colleaux:2017ibe,Colleaux:2019ckh,Borissova:2026krh} and the low dimensionality of spherically symmetric systems, there is essentially a theory for each static black hole.
 
Given the decomposition  of the Ricci scalar and Einstein tensor  in \eqref{DSS}, which are given by
\begin{eqnarray}
\label{Riccidecom}
R =R\left(\gamma \right)  +  R\left(\Omega \right) -  2 \, \frac{ D^2 r^n}{r^n} +n(n-1) \frac{ D r .  D r}{r^2} 
\end{eqnarray}
and
\begin{eqnarray}
\begin{split}
\label{Einsteindecom}
G_{ab} &=   - \frac{n(n-1)}{2}  \left(\frac{k-D r . D r}{r^2}\right) \gamma_{ab}  - \frac{n}{r}  \left( D_a D_b - \gamma_{ab} D^2 \right) r  \,,
\\ 
G_{\rho}^{\rho} = G_{\phi}^{\phi}&=   - \frac{(n-1) (n-2)}{2} \left(\frac{k-D r . D r}{r^2}\right)     +(n-1)   \frac{D^2 r}{r}  -\frac{1}{2}   R(\gamma)   \, ,
\end{split}
\end{eqnarray}
the theory \eqref{fullaction} uniquely follows applying the previous dictionary to the unique class of 2D scalar-tensor theories satisfying the following properties : 1. The 2D reduced Lagrangian consists of high-energy corrections coming from $d$-dimensional curvature invariants, which preserve the decomposition of the Ricci scalar \eqref{Riccidecom}, so it is of the form
\begin{eqnarray}
\sqrt{-\gamma} r^n \left( \psi_1\left(\mathcal{R}\right) R\left(\gamma \right)  + \psi_2\left(\mathcal{R}\right)  R\left(\Omega \right) +  \psi_3\left(\mathcal{R}\right)  \frac{ D^2 r^n}{r^n} + \psi_4\left(\mathcal{R}\right) \frac{ D r .  D r}{r^2} \right) \,; \label{LCdt}
\end{eqnarray}
2. Similarly for the 2D reduction of the field equations\footnote{Notice that, due to the scalar-tensor nature of the reduced theory and its 2D diffeomorphism invariance, the angular part of the metric field equation tensor, or equivalently the field equation associated with the scalar field $r$, is redundant compared to the field equation associated with the 2D metric.}, assumed to be of the form
\begin{eqnarray}
 \gamma_{ab}\left( \varphi_1\left(\mathcal{R}\right) \frac{k}{r^2} + \varphi_2\left(\mathcal{R}\right) \frac{Dr.Dr}{r^2} + \varphi_3\left(\mathcal{R}\right) \frac{D^2 r}{r} \right) + \varphi_4\left(\mathcal{R}\right) \frac{D_a D_b r}{r}  \;; \label{GCdt}
\end{eqnarray}
3. In even dimensions, the critical order scalar $n=2p$ contributes to the field equations \,; 
4. The 2D Dirichlet variational principle is well-defined. For more details about this short calculation, see \cite{Colleaux:2019ckh}. In particular, $\psi_3$ corresponds to a boundary term and is fixed from 4., while 1. and 2. imply $\left( r^{n-2}\psi_4 \right)=  d^2 \left(r^n \psi_1 \right)/ d r ^2$, meaning that the 2D dilaton gravity studied in \cite{Barenboim:2025fds, Barenboim:2025ckx} is a particular case of the 2D spherically symmetric sector of our theory. As we are interested in well-defined high-energy corrections in spacetimes \eqref{DSS}, we further require the functions $\psi$ to be expressible as power-series.

\subsubsection{Reduced action \& field equations}

In spacetimes (\ref{DSS}), we can integrate out the angular part of the Lagrangian, so that the action simply reduces to : 
\begin{eqnarray}
\label{2Daction}
 I  =\frac{V_{n,k}}{16 \pi G} \sum _{p=-1}^m \ell^{2p} \,k^{p} \, \left( \curv{I}_{p} +   \frac{ 2\epsilon \beta_{p}}{n-2p} \, \oint_{\partial \Sigma} dl \,  \sqrt{|h|} \, r^{n-2p}  \, K^{(2)}  \right)\,,
\end{eqnarray}
where
\begin{eqnarray}
 \curv{I}_{p} = \int_\Sigma d^{2}x \, \sqrt{-\gamma} \,  \left( \alpha_{p} \, k + \beta_{p} \left( (n-2p-1) D r . D r +  \frac{r^2}{n-2p}  \, R(\gamma)  \right) \right)r^{n-2p-2} \,.
\end{eqnarray}
Incidentally, this confirms that the 2D theory has a well-defined Dirichlet variational principle, as it yields the Gibbons–Hawking–York boundary term of Einstein-dilaton gravity in 2D.

Deriving the field equations associated with the 2D metric $\gamma$ and the scalar field $r$ shows that the non-vanishing components of the corrected Einstein tensor \eqref{FieldEq} for the metric \eqref{DSS} are given by  
\begin{eqnarray}
\begin{split}
\label{2DcovEOM}
&\mathcal{G}_{ab} = \sum _{p=-1}^m  \Big(-\frac{\Delta_\alpha}{2} \frac{k}{r^2}  \, \gamma_{ab} - \frac{\Delta_\beta}{r}  \left( D_a D_b - \gamma_{ab} D^2 \right) r 
+ \frac{\Delta_\beta}{2} (n-2p-1) \gamma_{ab} \frac{D r . D r}{r^2}   \Big)  \,, 
\\
&\mathcal{G}_{\rho}^{\rho} = \mathcal{G}_{\phi}^{\phi} =  -\frac{1}{2n} \sum _{p=-1}^m  \Phi_p \,,
\end{split}
\end{eqnarray}
where 
\begin{eqnarray}
\Phi_p =\Delta_\alpha (n-2p-2) \frac{k}{r^2}    -\Delta_\beta  \left((n-2p-1) \left( (n-2p-2) \frac{D r . D r}{r^2} + 2   \frac{D^2 r}{r} \right) -   R(\gamma) \right) \,,
\end{eqnarray}
and
\begin{eqnarray}
\Delta_\alpha =  \alpha_p \left( \ell^{2} \mathcal{R} \right)^p \,.
\end{eqnarray} 
Notice the contribution from the curvature corrections $p>1$ is trivial for flat topology $k=0$, meaning that the corresponding solutions are stealth. This may be seen as a weakness of this model, and of any non-polynomial gravity whose 2D decomposition belongs to the Einstein-dilaton subclass of Horndeski gravity.

The generalised Misner-Sharp quasi-local mass of this system is given by 
\begin{eqnarray}
M= \frac{V_{n,k}}{16 \pi G}   \sum_{p=-1}^m \left( \frac{\Delta_\alpha}{n-2p-1} \frac{k}{r^2} -\Delta_\beta \frac{Dr.Dr}{r^2} \right)r^{n+1}\,, \label{QLMass}
\end{eqnarray}
where the summation in odd dimensions is intended as a limit for the correction of order $p= (n-1)/2$. More precisely, it contributes to the mass as 
\begin{eqnarray}
  \lim_{p \to \frac{n-1}{2}}   \frac{\Delta_\alpha}{n-2p-1} \frac{k}{r^2} r^{n+1}= \alpha_{\frac{n-1}{2}} k^{\frac{n+1}{2}} \ell^{n-1} \log\left(r\right)\,,
\end{eqnarray}
modulo a divergent constant. Indeed, introducing the Kodama vector 
\begin{eqnarray}
K^a = - \varepsilon^{ab} D_b r \,,
\end{eqnarray}
where $\varepsilon$ is the 2D Levi-Civita tensor, enables to define a locally conserved current given by 
\begin{eqnarray}
J^a = -\frac{1}{8\pi G} \mathcal{G}^{ab} K_b =  - \frac{1}{r^n V_{n,k}} \varepsilon^{ab} D_b M \,,
\end{eqnarray}
so that $M$ is indeed a conserved quantity along $J$.

\subsection{Critical order theory and 2D Euler characteristic}\label{2dchi}

In even dimensions $d=2(s+1)$, the critical order correction $d-2= 2s=n=2p$, whose order equal the number of spacetime dimensions of $\mathcal{M}$, is ill-defined as it is, see \eqref{fullaction} and \eqref{Ib}. For instance, in four dimensions, this corresponds to the first correction, quadratic in the curvature. Consider instead the rescaled critical action 
\begin{eqnarray}
I_{\text{crit}}= \frac{\ell^n}{16\pi G } \left( \int_\mathcal{M} d^{\, n+2}x  \, \sqrt{-g}  \mathcal{R}^{s} \mathcal{S}_{s} +  2\epsilon \oint_\mathcal{\partial M} d^{\, n+1}x \,  \sqrt{|\lambda|}   \,   \mathcal{R} ^{s}   K \right)\,, \label{critaction}
\end{eqnarray}  
which can be obtained from \eqref{fullaction} and \eqref{Ib} as the dimensional limit $I_{\text{crit}}= \lim_{p\to s} (n-2p)\left( I_g + I_b \right)$ with $\beta_{s}=1$. Setting $V_{n,k} \ell^n k^{s}  =16 \pi G$, and $k\neq 0$, for otherwise it would be trivial, the 2D decomposition of this action reads
\begin{eqnarray}
 I_{\text{crit}}  =   \int_\Sigma d^{2}x \, \sqrt{-\gamma}   \, R(\gamma)    + 2\epsilon \, \oint_{\partial \Sigma} dl \,  \sqrt{|h|}  \, K^{(2)}   =4 \pi  \chi \,, \label{EulerCh}
\end{eqnarray}
where $\chi$ is the Euler characteristic of $\Sigma$, obtained through the Gauss-Bonnet theorem, assuming $\Sigma$ has smooth boundary. Therefore, when evaluated in \eqref{DSS}, this $d$-dimensional action is a  topological invariant of the 2D orbit space, so in particular it has identically vanishing field equations. This is why the factor $1/(n-2p)$ appearing in the action \eqref{fullaction} enables to obtain non-vanishing contribution to the field equations from the critical order theory, following the third requirement of \ref{Deftheory}. For more details on similar dimensional regularisations in the context of Lovelock-Lanczos gravity, see \cite{Colleaux:2020wfv}.

In order to obtain a well-defined $d$-dimensional regularised action with  non-trivial field equations coinciding with \eqref{2DcovEOM}, our prescription to make sense of the $p=n/2$ term of the series is the following.  Starting from the 2D action \eqref{2Daction} and integrating by part the Ricci scalar using that locally 
\begin{eqnarray}
R\left(\gamma\right) =-2 \epsilon D_a \left( n^a K^{(2)} \right), \label{RicciD}
\end{eqnarray}
 the boundary term identically vanishes and the dimensional factor in the bulk term disappears. It is then possible to take the limit $p \to n/2$ and re-express the action in terms of the Ricci scalar. This results in the regularised 2D Einstein-dilaton theory,
\begin{eqnarray}
\label{2Dactionreg}
 I_{\text{reg}}  =\frac{ \beta_{s}  k^{s} \ell^{n} V_{n,k}}{16 \pi G}\left( \int_\Sigma d^{2}x \, \sqrt{-\gamma} \,  \left(  \log \left( r\right)   R(\gamma) -\frac{Dr.Dr}{r^2}  \right) + 2 \epsilon  \oint_{\partial \Sigma} dl \,  \sqrt{|h|}  \log \left( r\right)  \, K^{(2)} \right)  .
\end{eqnarray}
The corresponding $d$-dimensional non-polynomial gravity action obtained through the dictionary \eqref{Dico} reads
\begin{eqnarray}
I_{\text{reg}}= \frac{\beta_{s} \ell^{n}}{16\pi G} \left( \int_\mathcal{M} d^{d}x   \sqrt{-g}  \,  \mathcal{R}^{s}   \, \mathscr{S}    -  \epsilon   \oint_\mathcal{\partial M} d^{d-1}x  \sqrt{|\lambda|}    \,  \mathcal{R} ^{s} \log\left(\frac{\mathcal{R}}{\mathcal{R}_0} \right)   K \right), \label{RegCritL}
\end{eqnarray}
with
\begin{eqnarray}
\mathscr{S} = -\frac{1}{2} \left( \frac{d}{ds}   + \log\left(\frac{\mathcal{R}}{\mathcal{R}_0} \right) \right) \mathcal{S}_{s}  \,,
\end{eqnarray}
where $\mathcal{R}_0$ is a reference curvature scale introduced for dimensional reasons. It does not affect the 2D dynamics, because the term involving it yield the critical action \eqref{critaction}.

Therefore, the property that non-analytic $d$-dimensional curvature invariants in the action become analytic for spacetimes \eqref{DSS}, which holds for the other invariants, is lost at critical order. However, this is in some sense the weakest non-analyticity, as it comes from the dimensional limit of a curvature monomial and its contribution to the field equations remains analytic.

\section{Black hole solutions and their properties}\label{BHsol}

In this section, we prove Birkhoff theorem and obtain the most general static black hole solutions of \eqref{fullaction}. Their temperature and entropy are derived and shown to satisfy the first law of black hole thermodynamics. It is shown that the regularised critical order correction \eqref{RegCritL} yields a quantum-like correction to the Newtonian potential \cite{Donoghue:1993eb} and a logarithmic correction to the entropy, which are well-known low-energy one-loop quantum gravity results \cite{Fursaev:1994te, Kaul:2000kf, Carlip:2000nv, Engle:2009vc, Sen:2012kpz}. Furthermore, we prove that these black holes can be made regular at the origin, replacing the classical singularity by an (A)dS regular core, assuming that the highest order of correction is larger than critical  $m>d/2$ and fixing a single coupling constant $\alpha_m$. The conditions on the couplings to have Schwarzschild-Tangherlini-(A)dS behaviour at infinity are also obtained.

\subsection{Birkhoff theorem and quasi-topological property}

Let us consider a diagonal gauge for \eqref{DSS},
\begin{eqnarray}
\label{Weylgauge}
d\Sigma^{\,2} = -a(t,r) b(t,r)^2 dt^2 + \frac{dr^2}{a(t,r)}\,,
\end{eqnarray}
where we have chosen the scalar field $r$ to be the radial coordinate, so that the field equations \eqref{FieldEq} in vacuum, whose components are given by \eqref{2DcovEOM}, can be expressed as
\begin{eqnarray}
 \dot{a} \sum_{p=-1}^m \Delta_\beta   = 0   \;,\;\;\;\;  b' \sum_{p=-1}^m \Delta_\beta  = 0  \;,\;\;\;\; \frac{k}{r^2} \sum_{p=-1}^m \Delta_\alpha -  \frac{1}{r^n}\left(   r^{n-1} \, a \sum_{p=-1}^m \Delta_\beta  \right)'=0 \,, \label{Eqfct}
\end{eqnarray}
where primes and dots are respectively derivatives with respect to the radius $r$ and time coordinate $t$. The second equation implies that the theory has single-function black hole solutions, which is the typical property of quasi-topological gravities. As a consequence of the first equation and reabsorbing $b(t)$ by a time rescaling, the solutions are time-independent and the theory satisfies Birkhoff theorem.

The most general solutions for an arbitrary number of coupling constants is given by, 
\begin{eqnarray}
\label{RRBHs}
\begin{split}
a(r) = \frac{1}{1+\frac{1}{n}\sum\limits_{p=1}^m \Delta_\beta} \Bigg(  -\frac{2}{n\left(n+1\right)} \Lambda r^{2}+k \left( 1 + \frac{1}{n} \sum\limits_{\substack{p\in \left[ 1 , m \right] \\ p\neq \frac{n-1}{2}}} \Delta_{\tilde{\alpha}}+ \frac{\Delta}{n}  \log\left(\frac{r}{r_0}\right) \right)-\mu  r^{1-n} \Bigg)\,,
\end{split}
\end{eqnarray}
where $r_0$ is a constant, $\mu$ can be written in terms of the quasi-local mass \eqref{QLMass}, which is time-independent due to Birkhoff theorem,  as
\begin{eqnarray}
\mu =\frac{16 \pi G M}{n V_{n,k}}  \,,\label{mu}
\end{eqnarray}
 and\footnote{Remark that $\Delta=0$ if $(n-1)/2$ is not an integer, i.e. in even dimensions or if $n-2m-1>0$.}
\begin{eqnarray}
\tilde{\alpha}_p = \frac{\alpha_{p}}{n-2p-1}   \;,\;\;\;\; \Delta_\alpha = \alpha_p \,k^p \left(\frac{\ell}{r}\right)^{2p}  \;,\;\;\;\; \Delta = \alpha_{\frac{n-1}{2}} \,k^{\frac{n-1}{2}} \left(\frac{\ell}{r}\right)^{n-1} . \label{Delta}
\end{eqnarray}
Therefore, except for the logarithmic term present in odd dimensions, the unique solutions of the theory (\ref{fullaction}) for spacetimes (\ref{DSS}) are all the rational single-function black holes with $g_{tt}$ linear in the mass $M$. The vacua of the theory, corresponding to $M=0$, are generically smooth small scale deformations of (A)dS or Minkowski geometry which reduce to maximal symmetry for $\ell=0$, i.e. for GR, or for $r\to \infty$, as we will see in \ref{Sec.vac}. However, it is possible to obtain a Minkowski vacuum considering the following fine-tuning between the two series of coupling constants, 
\begin{eqnarray}
\alpha_p = \left(n-2p-1 \right)   \beta_p \,,\label{M4Vac}
\end{eqnarray}
as well as $\Delta=0$ and $\Lambda=0$. In this case, restoring the mass yields the class of solutions
\begin{eqnarray}
a(r)= k - \frac{\mu}{r^{n-1} \left( 1+\frac{1}{n}\sum\limits_{p=1}^m \beta_p \,k^p \left(\frac{\ell}{r}\right)^{2p}  \right)}\,. \label{RBHM4}
\end{eqnarray}

\subsection{Logarithmic entropic correction and first law of black hole thermodynamics}

In appendix \ref{App}, we present a minisuperspace version of the Wald entropy for static black holes of the form \eqref{DSS} with horizon located at $r=r_H$, which reads 
 \begin{eqnarray}
 S= 4  \pi  \mathcal{A}_k \left. \frac{\partial L\big|}{\partial R\left(\gamma\right)}\right|_{r_H}    \,,
\end{eqnarray}
where 
\begin{eqnarray}
\mathcal{A}_k = r_H^n V_{n,k} 
\end{eqnarray}
is the area of the black hole horizon. Applying this formula to the reduced Lagrangian density $L$ given by \eqref{2Daction} for $n \neq 2p$ and to the regularised one \eqref{2Dactionreg} in even dimensions yields 
\begin{eqnarray}
S = \frac{\mathcal{A}_k}{4 G} \left( \sum\limits_{\substack{p\in \left[ 0 , m \right] \\ p\neq \frac{n}{2}}}  \frac{\beta_p}{n-2p}  \,k^p \left(\frac{\ell}{r_H}\right)^{2p} + \beta_{\frac{n}{2}} k^\frac{n}{2}\left(\frac{\ell}{r_H}\right)^{n} \log\left(\frac{r_H}{r_0} \right)  \right) \,,\label{SLog}
\end{eqnarray}
where $\beta_{n/2}$ vanishes in odd dimensions and $r_0$ is some arbitrary scale. Recall that $\beta_0=n$, so we recover the Bekenstein-Hawking entropy for GR, $p=0$. Furthermore, except for the logarithmic term, this entropy formula is the same as that of Lovelock black holes \cite{Whitt:1988ax,Maeda:2011ii}, and so of quasi-topological ones as well. Remark that the (even-dimensional) logarithmic correction is coming from the  regularised critical order scalar studied in \ref{2dchi}. Such correction has been predicted by various quantum gravity approaches \cite{Fursaev:1994te, Kaul:2000kf, Carlip:2000nv, Engle:2009vc, Sen:2012kpz}, and also appear in 4D regularisations of Gauss-Bonnet gravity \cite{Lu:2020iav, Kobayashi:2020wqy, Fernandes:2020nbq, Wei:2020poh}.
\medskip

Defining the temperature of the black holes in terms of the surface gravity $\kappa$ on the Killing horizon by
\begin{eqnarray}
T=\frac{\kappa}{2 \pi} = \frac{a'(r_H)}{4 \pi}=\frac{k}{4\pi r_H}  \left( \sum\limits_{p=0}^m \beta_p \,k^p \left(\frac{\ell}{r_H}\right)^{2p}\right)^{-1}\sum\limits_{p=-1}^m \alpha_p \,k^p \left(\frac{\ell}{r_H}\right)^{2p} \,,\label{TBH}
\end{eqnarray}
it is straightforward to show that the first law of black hole thermodynamics is satisfied 
\begin{eqnarray}
\delta M = T \delta S \,,
\end{eqnarray}
where $M$ is the quasi-local mass \eqref{QLMass} evaluated on the horizon. The heat capacity and free energy of these black holes can be found in Chap.III.D of \cite{Colleaux:2019ckh}.

\subsection{Regularity conditions \& asymptotic behaviour}

\subsubsection{Schwarzschild-Tangherlini-(A)dS}

In order for these black holes to have the usual Schwarzschild-Tangherlini-(A)dS behaviour for $r\to\infty$, we need to impose some relations between the coupling constants, given by 
\begin{eqnarray}
\alpha_p = \left(n-2p-1 \right) \Big(  \beta_p +\frac{2 \ell^2 \Lambda}{n^2(n+1)} \left( \beta_p  \beta_1 -n \beta_{p+1}  \right)  \Big) \,,\;\;\;\; \text{for all} \;\;\;\;1 \leq p \leq \frac{n-2}{2}\,. \label{ftinf}
\end{eqnarray}
For $\Lambda=0$, they are the same as those required for the theories to admit an exact Minkowski vacuum \eqref{M4Vac}. In four and five dimensions $n=2,3$, there is no relation to impose. The behaviour of the general solution (\ref{RRBHs}) at infinity then becomes : 
\begin{eqnarray}
\label{arinf}
\begin{split}
& a(r\to \infty) = -\frac{2 \Lambda}{n(n+1)} \, r^2  +  k\left( 1+ \frac{2 \beta_1 \ell^2 \Lambda}{n^2(n+1)} \right)-    \frac{\mu}{r^{\,d-3}}  \\
&+ \frac{k^{\frac{n+1}{2}} \ell^{n-1}}{n \, r^{\,d-3}} \left( \alpha_{\frac{n-1}{2}} \log\left(\frac{r}{r_0}\right)-\beta_{\frac{n-1}{2}} + \frac{2 \ell^2 \Lambda}{n^2(n+1)}\left( n \beta_{\frac{n+1}{2}}  - \beta_1\beta_{\frac{n-1}{2}}\right) \right)  +\dots\,.
\end{split} 
\end{eqnarray}
 In higher dimensions, the previous relations imply the presence of corrections in the action of the form $ \Lambda \, \ell^{2p+2} \, \mathcal{R}^{p}$, meaning that the cosmological constant appears in the corrections $1 \leq p \leq \frac{d-4}{2}$, if the couplings involved in the previous relations are non-vanishing.

\subsubsection{Quantum-like correction to Newton potential}

An important condition to satisfy for high energy corrections to Schwarzschild black hole is that they reproduce the one-loop quantum correction to the Newtonian potential, which in four dimensions is of the form $\left( G M \, \ell_P^2/r^3\right)$, where $\ell_P$ is the Planck length \cite{Donoghue:1993eb,dePaulaNetto:2021axj}. This correction is one of the few consistency check at one’s disposal in order to assess whether a modification of gravity is in accordance with what we know about low energy quantum gravity.

Interestingly, it is possible for the solution (\ref{RRBHs}). Indeed, restricting to four dimensions ($n=2$), spherical topology ($k=1$) and setting $\Lambda=0$, the behaviour of the general solution (\ref{RRBHs}) at infinity is given by 
\begin{eqnarray}
a\left(r\to \infty\right) = 1 - \frac{2G M}{r} - \frac{\alpha_1 + \beta_1}{2} \frac{\ell^2}{r^2} + \frac{\beta_1 \,G M \, \ell^2}{r^3}  + \dots\,.\label{QCorrNewt}
\end{eqnarray}
 Thus, providing that we identify the length scale $\ell$ with the Planck length $\ell=\ell_P$ and consider $\beta_1\neq 0$, such quantum-inspired correction is indeed present and comes from the dimensional regularisation \eqref{RegCritL} of the critical order scalar \eqref{critaction}, which is the leading high energy correction to 4D GR in the theory \eqref{fullaction}.  In even higher dimensions, setting all the couplings $\alpha_p = \beta_p=0$ for $1\leq p \neq (d-2)/2$ and $\Lambda=0$ for simplicity, \eqref{QCorrNewt} generalises to
\begin{eqnarray}
a\left(r\to \infty\right) =  1 - \frac{\mu}{r^{d-3}} - \left( \frac{\alpha_{\frac{d-2}{2}}+\beta_{\frac{d-2}{2}}}{r^{d-2}} - \frac{\beta_{\frac{d-2}{2}} \mu}{r^{2d-5}} \right) \frac{\ell^{d-2}}{d-2}+\dots\,,\label{QCorrNewtd}
\end{eqnarray}
which again agrees with the literature, see e.g. eq(48) of \cite{Garbarz:2008fr}. In order to have only these correction up to $O\left(r^{-2d+4}\right)$, it is required to set $\alpha_p=\left(n-2p-1\right) \beta_p$, for $p=(d-2)/2$, which is again the same class of conditions required for a Minkowski vacuum and Schwarzschild-Tangherlini-(A)dS asymptotic behaviour in higher dimensions.

\subsubsection{Massive \& massless (A)dS-cores}\label{secadscore}

It turns out that these black hole solutions, which are, in the sense of \ref{Deftheory}, the most minimal corrections to Schwarzschild geometry obtained from a pure gravity diffeomorphism invariant theory, can be made regular at the origin without the need of infinite tower of corrections ($m\to\infty$).

Notice first that the regularity is possible only when the maximal order of correction $\mathscr{O} = 2 (m+1)$ is greater than the critical order, given, as in Lovelock gravity, by the dimension $d$ of the spacetime, 
\begin{eqnarray}
\mathscr{O} > d \,,\;\;\;\;\;\; 2m > n\,,
\end{eqnarray}
 i.e. when the scalar $\mathcal{R}^{m} \left(\alpha_m \mathcal{R} + \beta_m \mathcal{S}_{(m)} \right)$ contains more derivatives than the number of dimensions. For instance, in $d=4,5$, it is necessary to consider at least $m=2$ cubic curvature corrections to cure the singularity. Taking into account this regularity condition and defining 
\begin{eqnarray}
\gamma_p = \beta_p - \frac{\alpha_p}{n-2p-1} \,,
\end{eqnarray}
assuming $a(0)=k$ implies the simple relation between $\alpha$ and $\beta$ : 
 \begin{eqnarray}
 \gamma_m=0\,.  \label{RegCdt}
 \end{eqnarray}
This fine-tuning belongs to the class of relations, $\gamma_p=0$ for some $p\in [1,m]$, needed to have both the quantum-like correction to Newton potential, Minkowski vacuum and Schwarzschild-Tangherlini behaviour at $r \to \infty$ in higher dimensional solutions with $\Lambda=0$.

Let us consider the smallest maximal order for which the singularity can be cured,  
\begin{eqnarray}
m = \left\lceil \frac{n}{2} \right\rceil \,.\label{critorder}
\end{eqnarray}
In even dimensions, it is possible to obtain an (A)dS-core only for vanishing mass $\mu=0$. Thus, in order to have regular black hole solutions of arbitrary mass in even dimensions, it is necessary to consider curvature corrections beyond the critical order. In odd dimensions, \eqref{critorder} is already above the critical order and assuming (A)dS-core implies \eqref{RegCdt} and the condition
\begin{eqnarray}
\alpha_{\frac{n-1}{2}}=0 \,,
\end{eqnarray}
corresponding to the absence of logarithmic term in the solution, so that
\begin{eqnarray}
a(r\to 0) = k +2 \left( \frac{\beta_{m-1} + n \mu k^{-m} \ell^{1-n}}{ \alpha_m} \right) \frac{r^2}{\ell^2} + \dots\,.\label{AdScoreOddCrit}
\end{eqnarray}
When the maximal order of correction is higher than the critical dimension,  
\begin{eqnarray}
\mathscr{O} > d \,,\;\;\;\;\;\; m > \left\lceil \frac{n}{2} \right\rceil \,,
\end{eqnarray}
the condition \eqref{RegCdt} is sufficient for the black hole to have an (A)dS-core, as it yields
\begin{eqnarray}
a(r\to 0) = k - \left( \frac{\gamma_{m-1} }{ \beta_m} \right) \frac{r^2}{\ell^2} + \dots\,.\label{AdScore}
\end{eqnarray}

Finally, remark that the black holes with Minkowski vacuum \eqref{RBHM4} satisfy \eqref{M4Vac}, i.e. $\gamma_p=0$. Therefore, they are automatically regular and admit a  Minkowski-core given by 
\begin{eqnarray}
a(r\to 0) = k - \frac{n \mu}{\beta_m k^m r^{n-1}}\left( \frac{r}{\ell} \right)^{2m}+ \dots\,.\label{AdScoreM4}
\end{eqnarray}
From the point of view of curvature invariants, this type of ``massive cores", appearing also in \eqref{AdScoreOddCrit}, implies that polynomial invariants in the Riemann tensor are either vanishing at $r=0$ or finite and proportional to the mass $\mu$. In some sense, this is not ideal because, if \eqref{fullaction} is interpreted as a toy model for quantum corrected black holes $\ell=\ell_P$, the curvature would be expected to be Planckian at $r=0$, i.e. independent from the mass and proportional to $1/\ell_P^2$. From this point of view, the (A)dS-cores of the black hole solutions should be ``massless". We will propose below a mechanism to resolve this issue, based on alternative vacua than flat space which nonetheless reduce to it when $\ell=0$.  

\section{Rational regular black holes}\label{RRBH}

In this section, we present the most general regular black hole solutions  of \eqref{fullaction}, study their smoothness and compare them with their QTG and Designer Lovelock Gravity cousins. Moreover, we classify their vacuum structure $M=0$ and present a class of theories admitting extremal black hole vacua satisfying Nernst's third law of thermodynamics.

\subsection{Most general regular solutions}

Combining the previous results together, the most general black hole solutions of \eqref{fullaction}, behaving asymptotically as Schwarzschild-Tangherlini-(A)dS and admitting a central (A)dS-core, are the single-function metric
\begin{eqnarray}
ds^2 = - a(r)dt^2 + \frac{dr^2}{a(r)} + r^2 d\Omega_{n,k}\,,\label{SFsol}
\end{eqnarray}
which are linear in the mass and rational functions of the radius, up to an odd-dimensional log term, 
\begin{eqnarray}
a(r)= k - \frac{\frac{2 \Lambda}{n\left(n+1\right)} \left(r^2 - \frac{k \ell^2}{n^2} \sum\limits_{p=1}^{\lfloor \frac{n-2}{2} \rfloor} \Delta_\zeta \right) + \mu r^{1-n}+  \frac{k}{n}\left(\bar{\Delta}-  \Delta \log\left(\frac{r}{r_0}\right) + \sum\limits_{p=\lceil \frac{n}{2} \rceil}^{m-1}  \Delta_\gamma \right)}{1+\frac{1}{n}\sum\limits_{p=1}^m \Delta_\beta} \,,\label{RBHSTAdS}
\end{eqnarray}
and satisfy $m > \frac{n}{2}$, as well as $\Delta=0$ when $m = \left\lceil \frac{n}{2} \right\rceil$ in odd dimensions, where 
\begin{eqnarray}
\zeta_p= \beta_p  \beta_1 -n \beta_{p+1} \,, \label{zeta}
\end{eqnarray}
and 
\begin{eqnarray}
\bar{\Delta} = \beta_{\frac{n-1}{2}} \,k^{\frac{n-1}{2}} \left(\frac{\ell}{r}\right)^{n-1} \,, \label{barDelta}
\end{eqnarray}
which vanishes in even dimensions, while the other deltas are given by \eqref{Delta}. They are obtained from the solutions \eqref{RRBHs} of the general theory \eqref{fullaction} from \eqref{RegCdt} and \eqref{ftinf}, which represents together
\begin{eqnarray}
1 +\left\lfloor\frac{d-4}{2}\right\rfloor
\end{eqnarray}
relations among the coupling constants. In order to prevent the appearance of curvature singularities at some finite radius $r\neq 0$, the denominator in \eqref{RBHSTAdS} should be non-vanishing for all $r$, i.e.
\begin{eqnarray}
\sum_{p=0}^{m} \beta_p k^p \left( \ell^{p} r^{m-p} \right)^2 >0 \,,
\end{eqnarray} 
with $\beta_0=n$, what implies some mild fine-tuning of the coupling constants $\beta$, which must satisfy some inequalities. For instance, in order to have both regular spherical black holes and regular AdS hyperbolic ones, with $k=-1$ and $\Lambda<0$, a sufficient condition is to have $\beta_p>0$ and $m$ even.

\subsubsection{Smoothness \& comparison with other pure gravity theories}\label{secsmoothness}

There has been some interesting new results regarding the regularity of differential curvature invariants \cite{Giacchini:2021pmr} and geodesic completeness \cite{Zhou:2022yio} in regular black holes with (A)dS-cores. In particular, it was shown that a lack of smoothness of such geometries around $r=0$, what happens when the function $a(r)$ is not even, produces some divergence of differential curvature invariants, roughly at the order of the first odd power of $r$ around the (A)dS-core.

 Setting the logarithmic term to zero, we see that all the massless geometries $\mu=0$ are smooth, the odd-dimensional solutions \eqref{RBHSTAdS} are smooth for any mass, and none of the even-dimensional ones are for $\mu\neq 0$.  However, in the latter case, the first odd power of $r$ in the expansion of the solutions around $r=0$,
\begin{eqnarray}
a(r\to 0 ) = \sum_{j=0}^{i} a_j r^{2j}  - \frac{n \mu}{\beta_m k^m \ell^{2m}} r^{2i+1} + \dots \label{EFTsmoothness}
\end{eqnarray}
 is related to the maximal order of correction  $m$ by
\begin{eqnarray}
i = m - \frac{n}{2} \,.
\end{eqnarray} 
Thus, if considered as an effective field theory, in the sense that describing higher and higher energies requires to take more and more corrections into account in the action, the smoothness is not a serious issue, as there is always a finite order of correction for which the divergence of a given differential curvature invariant can be avoided,  without requiring further fine-tuning than the single relation needed for the (A)dS-core. This is similar to what happens in Designer Lovelock Gravity \cite{Kunstatter:2015raa,Kunstatter:2015vxa} and its higher-dimensional QTG \cite{Bueno:2024dgm} and NPG \cite{Colleaux:2017ibe,Colleaux:2019ckh} formulations, where, although any truncation of the action yields singular black holes, an (A)dS-core appears in the non-perturbative limit for very mild fine-tuning \cite{Bueno:2024dgm}. However, given that the solutions of these theories do not have an (A)dS-core at finite $m$, the smoothness issue does not seem possible to address perturbatively (in $\ell$) in these theories, contrary to \eqref{fullaction} whose regular black holes satisfy \eqref{EFTsmoothness}.

It is quite important to emphasise that contrary to DLGs, the theory \eqref{fullaction} do not require an infinite sum of curvature invariants to avoid the black hole singularities, but just a few non-analytic invariants. While the $d\geq 5$ QTG \cite{Bueno:2024dgm} are analytic in the curvature, what is certainly preferable, this is arguably an advantage of the present model over other 4D non-polynomial gravities, the disadvantage being stealth planar $k=0$ solutions and absence of limiting curvature, meaning, for static (A)dS-core black holes, that $a(r)$ is not a regular mass-independent geometry in the limit of infinite mass $\mu \to \infty$. To compare our solutions with rational black holes with limiting curvature, see \cite{Frolov:2016pav}.

Interestingly, the condition to avoid the classical singularity, which is to consider corrections to GR containing more derivatives than the critical order, meaning sixth order invariants in 4D, has an interesting counterpart with \cite{Giacchini:2018gxp, Giacchini:2018wlf}, where it was reported that the Newtonian limit of polynomial higher order gravities becomes regular at the origin precisely when considering at least sixth order scalars. It would be interesting to generalise these results to higher-dimensions and see if the analogy with our $d$-dimensional theory still holds.

\subsubsection{Minimal four \& five dimensional theories}

Let us present the minimal four and five dimensional theories admitting regular black holes for spherical and hyperbolic horizon topologies, with non-massive (A)dS-cores. In four dimensions, considering the regularised critical theory, which is quadratic in curvature, to ensure the quantum-like correction \eqref{QCorrNewt} and logarithmic correction to the entropy \eqref{SLog}, the minimal action involves up to cubic invariants and reads
\begin{eqnarray}
I_{4D} = \frac{1}{16 \pi G}  \int d^4 x \sqrt{-g} \left(-2 \Lambda+  R +\left( \ell^{2}  \mathcal{R} \right) \left( \alpha_1 \mathcal{R} + \beta_1  \mathscr{S}\right) - \beta_{2} \left( \ell^{2}  \mathcal{R} \right)^{2} \left( 3 \mathcal{R}+\frac{\mathcal{S}_{2}}{2} \right) \right)\,,\label{4Dmin}
\end{eqnarray}
with the associated solution 
\begin{eqnarray}
a(r)= k - \frac{r^2 \Big( \frac{1}{3} r^4 \Lambda + \mu r + \frac{1}{2}\left(\alpha_1+ \beta_1 \right) k^2 \ell^2 \Big)}{r^4 +\frac{1}{2} k \left( \beta_1 r^2 + \beta_2 k \ell^2 \right)\ell^2 }\,,\label{4Dminsol}
\end{eqnarray}
which has non-singular polynomial Riemann invariants for $k=\pm 1$ considering
\begin{eqnarray}
\beta_2 > \frac{\beta_1^2}{8} \,, \label{ftkpm}
\end{eqnarray}
and massless (A)dS-core \eqref{AdScore} for 
\begin{eqnarray}
\alpha_1 + \beta_1 \neq 0\,.
\end{eqnarray}
In five dimensions, the minimal theory with an (A)dS-core and regular solutions for both $k=\pm 1$ is cubic in the curvature but has a massive (A)dS-core. Furthermore, it is impossible for the quartic theory to have regular solutions for both $k=\pm 1$. Thus, the minimal theory satisfying both conditions is quintic in curvature. For instance,
\begin{eqnarray}
I_{5D} = \frac{1}{16 \pi G} \int d^5 x \sqrt{-g} \left( -2\Lambda+ R +\alpha_3 \left( \ell^{2}  \mathcal{R} \right)^{3}  \mathcal{R} - \beta_{4} \left( \ell^{2}  \mathcal{R} \right)^{4} \left( 6 \mathcal{R}+\frac{\mathcal{S}_{4} }{5}\right)\right)\,, \label{5Dmin}
\end{eqnarray}
admits smooth black hole solutions given by
\begin{eqnarray}
a(r) = k - \frac{r^2 \Big(\frac{1}{6}r^8 \Lambda + \mu r^4 +\frac{1}{12}\alpha_3 k^4 \ell^6 \Big)}{r^8 +\frac{1}{3} \beta_4 k^4 \ell^8}\,,\label{5Dminsol}
\end{eqnarray}
with $\beta_4>0$.

\subsection{Vacuum structure}\label{Sec.vac}

Let us now investigate in more details what are the possible vacua $M=0$ of these theories, other than Minkowski spacetime \eqref{RBHM4} which we previously identified. As it is sometimes the case from quantum gravity inspired corrections to Schwarzschild geometry such as polymeric black holes, as well as black bounce geometries and wormholes  \cite{Taves:2014laa,Peltola:2008pa,Peltola:2009jm,Modesto:2008im,Visser:1997yn,Simpson:2018tsi,DAmbrosio:2018wgv}, the vacua of the theory \eqref{fullaction} are generically smooth deformations of maximally symmetric geometries, which are recovered asymptotically or for $\ell \to 0$.  More precisely, we will see that they can either be anisotropic horizonless spacetimes or regular smooth, possibly multi-horizons, black holes, in particular extremal ones. Although this is not standard, non-maximally-symmetric vacua, such as Lifshitz geometries, are not uncommon in field theories and can appear due to some symmetry breaking or higher-order gravitational corrections \cite{Dehghani:2010kd,Ayon-Beato:2010vyw, Maeda:2011ii}. Deformations of Minkowski geometry have also appeared in various quantum gravity approaches, see e.g. \cite{Snyder:1946qz, Seiberg:1999vs, Amelino-Camelia:2016gfx, Amelino-Camelia:2017utp, Amelino-Camelia:2025ask}, although some might suffer from fine-tuning issues to avoid large scale breaking of Lorentz invariance \cite{Collins:2004bp}.
 We restrict here to four dimensions $n=2$.
 
\subsubsection{Minimal unique vacuum theories}\label{MUVT}

Because they are not maximally symmetric, the vacua of \eqref{RBHSTAdS} with $\mu=0$ are generically different for each $m$, what is non-standard in gravitational theories. For instance, considering Lovelock-Lanczos gravity or a polynomial $F(R)$ gravity truncated at order $m$, there always exists a maximally symmetric vacuum,  although the specific dependence of the curvature on the coupling constants of the theory may change with $m$, the maximal order of corrections\footnote{Notice that although most Designer Lovelock Gravity have more than one maximally symmetric vacuum, with $\mu=0$ for some finite maximal order of correction $m$, these additional vacua are non-perturbative in the coupling constants of higher-order invariants and might be interpreted in the context of effective field theory as truncation artefacts, the only relevant vacuum in this context being the one connected to low-energy, see e.g. \cite{Platania:2020knd} for a similar argument regarding ghost degrees of freedom. However, it is interesting that the theory \eqref{fullaction} does not have such a proliferation of vacua.}.

Natural candidates to avoid this issue are the theories \eqref{fullaction} which admit a Minkowski vacuum \eqref{RBHM4}, defined by $\gamma_p=0$ and $\Lambda=0$, where in four dimensions
\begin{eqnarray}
\gamma_p = \beta_p - \frac{\alpha_p}{1-2p} \,.\label{gamma4d}
\end{eqnarray}
 However, we saw that the associated black hole solutions have massive (A)dS-cores  \eqref{AdScoreM4}, what is not ideal, as discussed in \ref{secadscore}. Furthermore, it is important to notice that non-polynomial gravities adapted to \eqref{DSS} are generically constructed from non-analytic curvature invariants in the Weyl tensor, making their field equations ill-defined when evaluated on conformally flat geometries, such as maximally symmetric spacetimes and FLRW cosmologies. Therefore, this class of metrics can only be obtained as limiting solutions, because the NPG are constructed to have analytic field equations in \eqref{DSS}, so that this issue disappears after the 2D reduction, see Chap.III.H of \cite{Colleaux:2019ckh} for more details.  This might be a mild issue, but as the theory \eqref{fullaction} offers the possibility to have exact non-conformally flat vacua, it is worth exploring this possibility. 

\medskip

Imposing the vacuum to be shared by all $m>1$ truncations\footnote{Notice that one could as well construct theories sharing the vacuum of higher-order theories, e.g. that of $m=3$. However, we study here the simplest class.} satisfying $\gamma_m =0$, which guarantees the presence of an (A)dS-core, implies that the solutions \eqref{RBHSTAdS} can be decomposed as
\begin{eqnarray}
a_m(r) = a_{\text{vac}}(r) - \frac{\mu}{r}\,  \left( 1+\frac{1}{2}\sum\limits_{p=1}^m  \beta_{p} \,k^p \left(\frac{\ell}{r}\right)^{2p} \right)^{-1} \,, \label{uniqueminvac}
\end{eqnarray}
where  $a_{\text{vac}}$ is the vacuum of the minimal correction, $m=2$, able to cure the singularity, corresponding to the cubic theory \eqref{4Dmin}.  From \eqref{4Dminsol}, it is given by
\begin{eqnarray}
a_{\text{vac}}(r)= \left(- \frac{\Lambda r^6}{3} + k r^4 - \frac{1}{2}\left( \alpha \, r^2- \delta \, k\, \ell^2 \right) \ell^2 \right) \left( r^4 + \frac{1}{2} \left(\beta \, k\,  r^2  + \delta \, \ell^2 \right)\ell^2 \right)^{-1}\,, \label{SharedVac}
\end{eqnarray}
where we consider $k\neq 0$ and in order to have smooth vacua for both horizon topologies, we assume
\begin{eqnarray}
0<\frac{\beta^2}{8} <\delta  \,. \label{RegHypVac}
\end{eqnarray}
By construction, the coupling constants of the lowest order theory $m=2$ satisfy
\begin{eqnarray}
\alpha_{1}= \alpha  \,,\;\;\;\; \beta_{1}= \beta  \,,\;\;\;\; \beta_{2}= \delta  \,,\;\;\;\; \text{for} \;\;\;\; m=2 \,. \label{m2theory}
\end{eqnarray}
 Comparing \eqref{uniqueminvac} with \eqref{RBHSTAdS}, we obtain the following system of equations,
\begin{eqnarray}
\begin{split}
0 &= \Lambda \left( \beta - \beta_1 \right) \,, \\
0 &= - \left(\alpha +\beta \right) + \gamma_1 + \frac{1}{3}\Lambda \ell^2  \left( \delta  - \beta_2  \right)  \,, \\
0 &= - \left( \alpha +\beta \right) \beta_1+ \beta \gamma_1 + 2 \gamma_2 - \frac{2}{3} \beta_3 \Lambda \ell^2   \,, \\
0 &= - \left( \alpha +\beta \right) \beta_{m-1} + \beta \gamma_{m-1} + \delta \gamma_{m-2} \,, \\
0 &= - \left( \alpha +\beta \right) \beta_m + \delta \gamma_{m-1}  \,, \\
\end{split}\label{recgamma1}
\end{eqnarray}
and for $4\leq j\leq m$,
\begin{eqnarray}
0 = - \left( \alpha +\beta \right) \beta_{j-2} + 2 \gamma_{j-1} + \beta \gamma_{j-2} + \delta \gamma_{j-3} -\frac{2}{3} \beta_j \Lambda \ell^2 \,, \label{recgamma2}
\end{eqnarray}
which can be solved for $\gamma_p$ and the two highest order $\beta_m$ and $\beta_{m-1}$. Furthermore,  if the cosmological constant is non-vanishing, the coupling constant $\beta_1$ is fixed for all theories. Thus, for respectively $\Lambda\neq 0$ and $\Lambda = 0$, the theory of highest order $m$ has $m-3$ and $m-2$ free coupling constants which do not characterise the vacuum, as $\alpha, \beta, \delta$ and $\Lambda$ do, but only appear in the mass function in \eqref{uniqueminvac}.

In order to solve this system, we introduce without loss of generality an auxiliary set of coupling constants $\sigma$ such that the solution takes the form
\begin{eqnarray}
a_m(r) = a_{\text{vac}}(r) - \frac{\mu \, \ell^4}{r}  \left( r^4 + \frac{1}{2} \left(\beta \, k\,  r^2  + \delta \, \ell^2 \right)\ell^2 \right)^{-1}\left( \sum_{p=2}^m  \sigma_p \left( \frac{k \ell^2}{r^2}\right)^p  \right)^{-1}\,.
\end{eqnarray}
Comparing with \eqref{uniqueminvac}, we obtain
\begin{eqnarray}
\beta_p =  \;\; \left\{
  \begin{array}{@{}ll@{}}
 \;\;  \delta \sigma_p + \beta \sigma_{p+1} + 2 \sigma_{p+2}  \text{\;\;\;, \;\;\;\;  for \;\;\;\; $1 \leq p \leq m-2$,}   
 \\ 
 \;\; \delta \sigma_{m-1} + \beta \sigma_{m}  \text{\;\;\;, \;\;\;\; \;\;\;\; \;\;\;\; \,  \phantom{for} \;\;\;\;  $p=m-1$,}  
 \\ 
 \;\; \delta \sigma_{m}  \text{\;\;\;, \;\;\;\;\;\;\;\; \;\;\;\; \;\;\;\; \;\;\;\;\;\;\;\; \;\;\;    \phantom{for} \;\;\;\;  $p=m$,} 
  \end{array}\right. \label{UVT1}
\end{eqnarray}
This expression contains the same recurrence appearing in \eqref{recgamma2}, so that the latter can be simply solved, together with \eqref{recgamma1}. This yields 
\begin{eqnarray}
\gamma_p =  \;\; \left\{
  \begin{array}{@{}ll@{}}
 \;\;  \left( \alpha+\beta\right) \sigma_{p+1} + \frac{2}{3}\Lambda\ell^2  \; \sigma_{p+3}  \text{\;\;\;, \;\;\;\;  for \;\;\;\; $1 \leq p \leq m-3$,}
 \\ 
 \;\; \left( \alpha+\beta\right) \sigma_{m-1}  \text{\;\;\;, \;\;\;\;\;\;\;\; \;\;\;\;\;\;\;\; \;\;\;\; \,  \phantom{for} \;\;\;\;  $p=m-2$,} 
 \\ 
 \;\; \left( \alpha+\beta\right) \sigma_{m}  \text{\;\;\;, \;\;\;\; \;\;\;\; \;\;\;\; \;\;\;\;\;\;\;\;\;\;\;\;  \phantom{for} \;\;\;\;  $p=m-1$,} 
 \\ 
 \;\; 0   \text{\;\;\;, \;\;\;\; \;\;\;\; \;\;\;\; \, \;\;\;\;\;\;\;\;\;\;\;\; \;\;\;\;\;\;\;\;\;\;\;\;\; \phantom{for} \;\;\;\;  $p=m$,} 
  \end{array}\right. \label{UVT2}
\end{eqnarray}
where $\sigma_{1} = 0$, $\sigma_2 = 1$ and 
\begin{eqnarray}
\Lambda \sigma_3 = 0  \,,
\end{eqnarray}
meaning that $\sigma_3$ must vanish for all $m$ when $\Lambda\neq 0$ and is arbitrary otherwise.

\subsubsection{Typology}

Let us briefly comment on the possible vacua \eqref{SharedVac}. The vacua being smooth and symmetric under $r\to -r$, we consider the $r>0$ region. For $\Lambda>0$ and $k=1$, they possess a cosmological horizon located approximatively at
\begin{eqnarray}
r_{\text{cosmo}} \approx \sqrt{\frac{3}{\Lambda} \left( 1 + \frac{1}{6} \beta \Lambda \ell^2 \right)}\,, \label{rcosmo}
\end{eqnarray}
using \eqref{arinf}. There are two main types of vacua. The firsts are anisotropic, horizonless and interpolate smoothly between two (A)dS$_4$ geometries, a large scale one with cosmological constant $\Lambda$, and a small scale one with effective cosmological constant given by 
\begin{eqnarray}
\Lambda_{\text{eff}} =  \frac{3 \left(\alpha + \beta \right)}{\delta \ell^2}\,. \label{Lambdaeff}
\end{eqnarray}
 Otherwise, the vacua can have either one non-degenerate\footnote{One-horizon (massless) black hole vacua are only possible for $\Lambda<0$ and $k=-1$. They also appear in Einstein-Gauss-Bonnet gravity \cite{Cai:2001dz}.}, two separate horizons in the region $r>0$, or a degenerate one.  There are two main issues with non-degenerate black hole vacua. When two horizons are present, the inner one is a Cauchy horizon and suffers from instabilities, such as \cite{Maeda:2005yd} and mass-inflation \cite{Poisson:1989zz}. Furthermore, whether one or two non-degenerate horizons are present, the outer one will have non-vanishing universal temperature, i.e. only depending on the coupling constants of the theory. However, if a finite temperature is emitted by the vacuum, it might produce a back-reaction, making it unstable.  Regarding massless extremal black hole vacua, although it is known that exact extremal geometries suffer from the Aretakis instability \cite{Aretakis:2012ei}, any perturbation should also break the extremality, so that Aretakis charges should be only approximately conserved, leading to some significant but transient effects at the horizon \cite{Murata:2013daa}. The vacuum should then be recovered via evaporation in an infinite amount of time, if the usual third law of black hole thermodynamics is satisfied, although classical counterexamples are known \cite{Kehle:2022uvc}. In this sense, gravitational theories with extremal vacua might be akin to systems relaxing to their ground states only asymptotically, such as a damped oscillator.

\subsubsection{Extremal black hole vacua \& Nernst's third law of thermodynamics}\label{secexvac}

Within this bestiary of vacua, we will be interested here in those which admit an extremal spherical horizon. Within spherical symmetry, these vacua are arguably the most symmetric ones which describe small-scale deformations of (A)dS$_4$ and M$_4$, as they smoothly interpolate between these asymptotically, a product of maximally symmetric geometries describing the near-horizon geometry, and an (A)dS$_4$ core with effective cosmological constant given by \eqref{Lambdaeff}. The Nariai geometry, given by the product dS$_2 \times$ S$^2$, requires a positive cosmological constant and would not exist for $\Lambda=0$, so a Bertotti-Robinson \cite{Bertotti:1959pf,Robinson:1959ev} small scale AdS$_2 \times $S$^2$ structure should be preferred, as our analysis will confirm. This endows the vacuum with an AdS region without the need for a negative cosmological constant. For more details on near-horizon geometries and their application to the AdS/CFT correspondence, see e.g. \cite{Moitra:2018jqs}.

\medskip

 Without loss of generality, we can fix the location of the extremal horizon at $r=\ell$ by rescaling of $\ell$ and $\beta$. The corresponding theories are then obtained setting 
\begin{eqnarray}
\alpha= 4 - 2  \Lambda \ell^2 \;,\;\;\;\;\;\;\;\;\;\; \delta = 2 - \frac{4}{3} \Lambda \ell^2\,,  \label{ExtremalFT}
\end{eqnarray}
so that \eqref{SharedVac} reduces to 
\begin{eqnarray}
a_{\text{vac}}(r)= \left(r^2 - k \ell^2 \right)^2 \left( r^4 +  \ell^4 - \frac{2}{3} \Lambda \ell^6 + \frac{k}{2} \beta r^2 \ell^2\right)^{-1} \left( k - \frac{\Lambda r^2}{3} - \frac{2 k  \Lambda  \ell^2}{3} \right)\,. \label{SharedVacExtr}
\end{eqnarray}
 The condition \eqref{RegHypVac} required to have a smooth $k=-1$ vacuum becomes
\begin{eqnarray}
\frac{2\Lambda }{3} < \frac{1}{\ell^2} \left(1 - \frac{\beta^2}{16} \right)\,, \label{ineqtopex}
\end{eqnarray}  
which is a very weak constraint assuming small $\Lambda$ and $\ell$ and reduces to $-4 <\beta < 4$
for vanishing cosmological constant.  Remark that the $k=-1$ vacuum is  cosmological for $\Lambda\geq 0$, or possesses a pair of horizons, a dS-core and is locally asymptotically AdS$_4$ for $\Lambda<0$, as usual for this kind of solutions. As mentioned previously, vacua with non-extremal horizons should be unstable. From this point of view, $\Lambda\geq 0$ is favoured in these models.

The near-horizon geometry of \eqref{SharedVac} can be found from the transformation $t\to t/\epsilon$, $r\to \ell + r \epsilon  $, followed by the limit $\epsilon\to 0$, which gives: 
\begin{eqnarray}
ds^2 = - \left( \frac{r}{\ell_0} \right)^2 dt^2 + \left(\frac{\ell_0 }{r} \right)^2 dr^2 +   \ell^2 \, \left( d\theta^2 + \sin^2 \theta d\phi^2 \right)\,,\label{Robert}
\end{eqnarray}
where 
\begin{eqnarray}
\ell_0 = \ell \, \sqrt{\frac{1}{2} \left(1  - \frac{1}{3} \Lambda \ell^2 + \frac{\beta}{4} \right) \left(1- \Lambda\ell^2\right)^{-1}} \,.
\end{eqnarray}
Thus, the degenerate horizon is of Bertotti-Robinson type for $\ell_0^2>0$. For asymptotically AdS$_4$ and M$_4$ solutions, $\Lambda\leq 0$, this is automatically satisfied requiring a smooth hyperbolic vacuum \eqref{ineqtopex}. For positive cosmological constant, this implies
\begin{eqnarray}
0<\Lambda < \frac{1}{\ell^2} \,, \label{CdtBRdS}
\end{eqnarray}
together with \eqref{ineqtopex}. This means that our choice of vacuum is the more physical one, as cosmological Narai-type of degenerate horizon would require a very large cosmological constant $\Lambda > 1 / \ell^2$. As long as 
\begin{eqnarray}
\ell \neq \ell_0 \,, \label{ellell0}
\end{eqnarray}
the geometry is not conformally flat, and is a genuine solution of our theory, because, as we said in the second paragraph of \ref{MUVT}, conformally flat geometries can only be obtained as limiting solutions of non-polynomial gravities, because they are generically non-analytic in the Weyl tensor. The previous condition is automatically satisfied for $\Lambda\leq 0$ if we demand smooth hyperbolic vacuum \eqref{ineqtopex}. For $\Lambda>0$, requiring \eqref{ineqtopex} and \eqref{CdtBRdS} reduces \eqref{ellell0} to 
\begin{eqnarray}
\beta \leq - \frac{12}{5} \,,\;\;\;\;\;\;\text{or} \;\;\;\;\;\;\;\; \beta \neq  4 \left( 1 - \frac{5}{3} \Lambda \ell^2 \right)\,.
\end{eqnarray}
Given that the value and sign of the coefficient controlling the quantum correction to Newton's potential is dictated by the nature and number of fields, the second condition might be preferred.

\medskip

The extremal ground states \eqref{SharedVacExtr} with $k=1$ have thermodynamical properties which are very different from the usual extremal states of (regular or not) two-horizons black holes. Indeed, consider the corresponding massive solutions \eqref{uniqueminvac} with an outer horizon located at some radius $r_H$. As the mass of the black hole decreases via evaporation, $r_H$ tends to the universal coupling constant $\ell$ and the temperature of the black hole vanishes, by definition. Moreover, as the location of the horizon is dictated by the theory, the entropy of the theories \eqref{SLog} tends to a universal constant given by
\begin{eqnarray}
S_{\text{vac}} = \frac{\pi \ell^2}{G} \left(1 +\beta_1 \log\left(\frac{\ell}{r_0}\right) +\frac{1}{2} \sum\limits_{p=2}^m  \frac{\beta_p}{1-p}  \right)\,. \label{Svac}
\end{eqnarray}
Due to the presence of the logarithmic correction, it is possible to set a reference of zero entropy $S_{vac}=0$ for the vacuum of each theories fixing the arbitrary scale 
\begin{eqnarray}
r_0= \ell \, \exp\left( \frac{1}{\beta_1} \left(1 + \frac{1}{2}\sum\limits_{p=2}^m \frac{\beta_p}{1-p}\right)\right)  \,. \label{r0logS0}
\end{eqnarray}
Therefore, even when generalized to the charged and rotating case, there will always be a zero-temperature state reached for $M=Q=J=0$, associated with a vanishing entropy. Thus, the existence of such extremal vacuum state provides a concrete realization of Nernst's third law of thermodynamics, stating that the entropy of a system must go to zero or to a universal constant as its temperature $T$ goes to zero  \cite{Wald:1997qp}. Furthermore, this provides a physical interpretation of the critical theory \eqref{RegCritL} responsible for the logarithmic correction, as it implements the idea that the entropy should be defined up to a constant, meaning that there should be an arbitrary scale, here $r_0$, able to ``normalise" it to zero\footnote{However, even when the log correction is absent $\beta_1=0$, the entropy reaches a universal constant value at zero temperature \eqref{Svac} and Nernst's law is satisfied. In order to obtain a set of theories with vanishing entropy for their common vacuum in this case as well, it is then required to impose a fine-tuning on $\beta_{m-2}$, similar to \eqref{UVT1}.}. Contrary to usual arguments in favour of a zero entropy for extremal black holes \cite{Hawking:1994ii,Teitelboim:1994az, Mitra:1996xi, Edery:2010cx}, this does not require a discontinuity in the entropy or an alternative treatment compared to the non-extremal case. However, when they exist, any other extremal black holes with $M\neq 0$ and $Q\neq 0$ are subject to the usual dilemma of having zero entropy at the price of a discontinuity, or a non-vanishing one. Remark that asymptotically Lifshitz black holes satisfying Nernst third law exist and their thermodynamics has been recently compared to that of Bose gas, see \cite{Arefeva:2024wqb} and reference therein. 

\medskip

A possible interpretation of these vacua \eqref{SharedVac} is given as follows. Consider Bronstein's semi-classical argument in which assuming the existence of relativistic black holes together with the uncertainty principle implies the existence of a minimal resolution scale, given by a Planck size black hole horizon\footnote{Indeed, the energy required to probe a region of size $\ell_P$ should produce a Planck size black hole. It is not possible to probe smaller distances because adding more energy would make the black hole grow, while sending less energy would not suffice to probe that scale. This is the usual relation between UV/IR mixing and minimal resolution scale in gravity \cite{Nortier:2025gmc, Dvali:2010ue, Castellano:2021mmx}.}, see e.g. \cite{Rovelli:2014ssa}. Let us assume this scale can be described by some effective (semi-)classical geometry where localization is lost for test particles. If such black hole should evaporate, smaller scales could be resolved, so it should have instead vanishing temperature and be extremal. If such Planck size extremal black hole should have vanishing entropy, then interpreting the entropy of black holes in terms of the information they hide to the asymptotic region \cite{Bekenstein:1973ur} might suggest that such Planck size black hole should be empty, meaning with zero mass $M=0$, for otherwise it would hide some microscopic details associated with that mass.  From this point of view, such extremal horizon is more akin to the cosmological horizon rather than those of massive black holes, as it is not present due to some physical formation mechanism, but due to the existence of a minimal resolution scale, similar to the maximal one set by the cosmological constant. Thus, considering a spherically symmetric system centred around a point of a macroscopically empty space and zooming in might result in a deformation of Minkowski spacetime containing an extremal black hole of size $\ell_P$ with zero entropy and vanishing mass, as described by \eqref{SharedVacExtr}.

\section{Infinite towers of curvature corrections}\label{Sec.inf}

In this section, we consider infinite towers of corrections to the Einstein-Hilbert action, $m\to\infty$, to address two issues arising from the regular black holes \eqref{RBHSTAdS} and their charged generalisations \eqref{RBHSTRNAdS}. Firstly, the even-dimensional ones are not exactly smooth, as discussed in \eqref{secsmoothness}, so we construct theories admitting exact smooth black holes. However, we show that their charged counterparts lose their smoothness and that more generally non-perturbative ($m\to\infty$) solutions can be interpreted in two different ways, resulting in distinct spacetimes.  Secondly, given that the spherical regular black holes \eqref{RBHSTAdS} have an (A)dS-core, they necessarily possess an inner Cauchy horizon for finite $m$,  what makes them unstable against the mass inflation instability \cite{Poisson:1989vv, Poisson:1989zz, Ori:1991zz}, which is controlled by the value of the inner horizon surface gravity. This is the essential conclusion we will be interested in, as our aim is not to model differently the phenomenon, but to propose two different mechanisms to avoid or tame the mass inflation instability within the theory \eqref{fullaction}. The first is simply to construct non-perturbative black hole solutions with a single horizon. The second is based on the theories \eqref{UVT1}, \eqref{UVT2} which admit an extremal black hole vacuum. It is shown that as the number of curvature corrections $m \to \infty$, the inner horizon becomes extremal, so that its surface gravity vanishes. This idea of using inner-extremal black hole to cure mass inflation has been proposed in \cite{Colleaux:2019ckh} and has recently been studied in \cite{Carballo-Rubio:2022aa, Franzin:2022aa, McMaken:2023aa, DiFilippo:2024mwm}.

 It is important to emphasize that resolving the mass inflation issue for neutral and non-rotating regular black hole is interesting but nonetheless very insufficient : in order to cure the singularity, one invokes the presence of a central (A)dS core, but in return, it requires the presence of an inner horizon which is not present in the classical case, and is unstable. Therefore resolving mass inflation for neutral non-rotating regular black holes is just a necessary condition for the (A)dS-core regularity to be a solid proposal to cure BH singularities.
The real issue with mass inflation arises in the charged and/or rotating cases, because the classical solutions do suffer from it. For this reason, after seeing how to cure the mass inflation of some neutral black holes of the present model, we will also consider the charged case, and see that the instability can still arise, but can also be avoided when specific bounds relating the charge and mass are satisfied, enabling a stable static black hole regime to exist.

\subsection{Smooth and non-perturbative black holes}\label{sec.smoothminf}

Let us start discussing the possibility and interest of constructing smooth black holes from \eqref{fullaction}, and to consider non-perturbative corrections $m\to \infty$ compared to large but finite ones. Recall that the odd-dimensional (A)dS-core black holes \eqref{RBHSTAdS} are already smooth for finite-order corrections and without fine-tuning. Furthermore, this smoothness is preserved when these solutions are charged \eqref{RBHSTRNAdS}. While even-dimensional black holes \eqref{RBHSTAdS} are not smooth for any finite order of corrections, they become smooth in the limit $m\to\infty$ without any fine-tuning \eqref{EFTsmoothness}. Moreover, the contribution from the charge is even in the radius \eqref{RBHSTRNAdS}, so it preserves this property.

Let us see some examples of infinitely fine-tuned smooth four-dimensional spherical symmetry $k=1$ solutions and consider Minkowski vacuum for simplicity, so we restrict to the theories \eqref{M4Vac} with $m\to \infty$ and $\Lambda=0$ with the solutions  \eqref{RBHM4}. The following theory admits a smooth Bardeen-like regular black hole, recently studied in \cite{Barenboim:2025ckx},
\begin{eqnarray}
 \beta_p = 6 (-2)^{-p} \frac{(2p-5)!!}{p!}\,, \;\;\;\;\;\;\;\; a(r) = 1 - \frac{\mu r^2}{\left( r^2 + \ell^2 \right)^{\frac{3}{2}}} \,, \label{Bardeen}
\end{eqnarray}
with logarithmic correction to the entropy and quantum-like correction to Newton potential, while the following theory yields a smooth regular black hole without these corrections, 
\begin{eqnarray}
 \beta_{3p-1}= \beta_{3p-2} = 0 \,, \;\;\;\;\;\;\;\;  \beta_{3p} =    (-2)^{1-p} \frac{(2p-3)!!}{p!} \,,\;\;\;\;\;\;\;\;  a(r) = 1 - \frac{\mu r^2}{\sqrt{r^6 + \ell^6}}\,. \label{smooth2}
\end{eqnarray}
The first interesting property of these solutions is that, when charged using the replacement \eqref{NeutralToChargedMax}, they lose their smoothness. Indeed, exact smoothness requires to fine-tune the couplings so that the resulting solution is even $a(r)=a(-r)$, but the charged replacement is given by  
\begin{eqnarray}
\mu \longrightarrow  \mu -\frac{q^2}{r}\,, \label{NeutralToChargedMax4D}
\end{eqnarray}
in four dimensions. Furthermore, while the addition of charge does not remove the (A)dS-core of black holes with finite-order corrections, as we saw in \ref{RobRBHQ}, in particular in four dimensions, it might do so for infinitely fine-tuned solutions. E.g. the two previous solutions become singular when charged, 
\begin{eqnarray}
a(r \to 0) = 1 + \frac{q^2 r}{\ell^3} - \frac{\mu r^2}{\ell^3} + \dots\,.\label{adssmoothy}
\end{eqnarray} 
It is of course possible to consider more general smooth solutions, such as 
\begin{eqnarray}
a(r) = 1 - \frac{\mu r^{2s}}{\left( r^{2u} + \ell^{2u} \right)^{\frac{1+2s}{2u}}} \,,\label{smoothmetric}
\end{eqnarray}
which reduces to the previous solutions for respectively $u=s=1$ and $u=3$, $s=1$, so that the associated charged black holes are regular for $s\geq 2$. However, this shows that the smoothness of infinitely fine-tuned solutions  is much more fragile than that of the perturbative theories in $\ell$ \eqref{fullaction}, as above a certain maximal order $m$, a Maxwell field cannot break the (A)dS-core \ref{RobRBHQ}.  Similar lose of regularity have been reported recently in \cite{Carballo-Rubio:2026mvj}. A way to resolve this issue is to consider non-perturbative corrections to the electric field, such as \eqref{Im} with $v\to \infty$, which preserve the smoothness when the electric potential is a regular even function of $r$. For instance, choosing
\begin{eqnarray}
\sum _{p=0}^\infty \xi_p \left( \ell^{2} \mathcal{R}\right)^p = \frac{\left(1+ \ell^{2} \mathcal{R}\right)^{5/2}}{1-2\ell^{2} \mathcal{R}}  =1 + \frac{9}{2} \ell^{2} \mathcal{R} + \dots 
\end{eqnarray}
yields the smooth electric potential
\begin{eqnarray}
\varphi(r) = \frac{Q}{4\pi} \frac{r^2}{\left( r^2 + \ell^2 \right)^{\frac{3}{2}}}\,,
\end{eqnarray}
which in turn yields smooth charged black holes using  \eqref{NeutralToCharged}. Although non-minimal couplings should be expected from quantum corrections, this mechanism to obtain smooth black holes requires such interactions to be present and (infinitely) fine-tuned for each matter sources.

Finally, notice that for arbitrary large but finite order $m$, the  black holes of the theories \eqref{Bardeen} and \eqref{smooth2} possess a Minkowski-core \eqref{AdScoreM4} rather than \eqref{adssmoothy}. Thus, for series with finite radius of convergence, the two interpretations of the theory, whether $1/m \ll 1$ or $m\to\infty$, yield completely different black holes at small scales. As any four-dimensional purely metric correction to General Relativity has necessarily ghost degrees of freedom, owing to their higher-order field equations \cite{Lovelock:1969vyr} and insufficient degeneracy \cite{Crisostomi:2017ugk}, these models are generically better interpreted as perturbative corrections with a given regime of validity rather than non-perturbative ones, see e.g. the discussions in \cite{DeFelice:2023vmj,Bueno:2023jtc}. In our approach, this is indicated by the robustness of the regularity for large $m$ compared to its fragility in the non-perturbative limit $m \to \infty$.

\subsection{Quasi-regular single-horizon black holes}

Although the non-perturbative limits should be considered with care and are not particularly suited to ensure regular black holes with robust smoothness, as we just saw, they can nonetheless be used to cure the mass inflation instability. The simplest way to do so is to consider static black holes without inner horizons, which are akin to wormhole and black bounce geometries \cite{Visser:1997yn,Simpson:2018tsi, DAmbrosio:2018wgv}.

As a concrete example, we consider the one-parameter family of non-perturbative theories, $m\to\infty$, with Minkowski vacuum \eqref{M4Vac} and \eqref{RBHM4}, given by 
\begin{eqnarray}
 \beta_p= -\frac{n}{3} \, 2^{-p}\,  (-1)^{s+n}  \frac{(2(s+n)+1)!! (2(p-s-n)-3)!!}{p!} \,,\;\;\;\;\;\; p\geq 1 \,, \label{BBtheory}
\end{eqnarray}
where $s\geq 0$ is an integer controlling the smoothness of the solutions. With this choice of couplings, the Lagrangian can be evaluated in closed form in terms of beta functions. For instance, the high energy corrections corresponding to the first term in the action \eqref{fullaction} read
\begin{eqnarray}
\sum_{p=1}^\infty \alpha_p \ell^{2p}\, \mathcal{R}^{p+1} = \frac{n}{3}  \left( 1 - n - \left(1 - n - (2+2s+n) \mathcal{R}\ell^2  \right) \left( 1-  \mathcal{R} \ell^2\right)^{-\frac{1}{2}+s+n} \right)\mathcal{R}\,.
\end{eqnarray}
Therefore, the Lagrangian is real only when the following bound is satisfied,
\begin{eqnarray}
\mathcal{R} \leq \frac{1}{\ell^2}\,,
\end{eqnarray}
providing some kind of built-in limiting curvature for the theory, given that the high energy corrections in the action \eqref{fullaction} are controlled by this curvature invariant  \eqref{RS}.

The unique charged static black hole solutions of this theory are given by 
\begin{eqnarray}
a(r) = k - 3 \left( \frac{\mu}{r^{n-1}}  - \frac{q^2}{r^{2(n-1)}}\right) \left(2+ \left(1- \frac{k \ell^2}{r^2}\right)^{\frac{1}{2}+s+n} \right)^{-1}\,, \label{aBB}
\end{eqnarray}
reduces to eq(209) of \cite{Colleaux:2019ckh} for $n=2$ and $s=0$ and are designed to behave asymptotically as Schwarzschild-Tangherlini-Reissner–Nordstr\"{o}m spacetime, to admit the quantum-like and logarithmic corrections to Newton's potential and to the entropy, and most importantly, to be real and regular only within the range 
\begin{eqnarray}
r\geq \ell \,, \label{r>ell}
\end{eqnarray}
for spherical horizon topology $k=1$, while being a standard hyperbolic regular solution for $k=-1$. This is similar to Modesto's semi-polymeric black hole \cite{Modesto:2008im}, which also possesses a small scale region where the metric becomes complex in Schwarzschild gauge. Notice that one-horizon regular black holes can also be obtained requiring a small scale Euclidean  region  in Schwarzschild gauge, as it happens for the D'Ambrosio-Rovelli \cite{DAmbrosio:2018wgv} and Visser-Hochberg-Simpson \cite{Visser:1997yn,Simpson:2018tsi} black bounces. This was shown in Chap.V of \cite{Colleaux:2019ckh}, where we also obtained 4D non-polynomial gravity theories admitting these three geometries and their charged generalisations as unique solutions.
\medskip

A brief analysis of the solutions shows that the spherical neutral solutions $q=0$ have at most one horizon and that the components of the $d$-dimensional Riemann tensor \eqref{Riemcomp} at $r=\ell$ are finite. Thus, these geometries avoid by construction the mass inflation instability while preserving the regularity of curvature invariants, contrary to respectively (A)dS-core and Schwarzschild-Tangherlini black holes. The presence of horizon is determined from $a(\ell)$, the spacetimes having one horizon when $a(\ell)\leq 0$ and none otherwise.

However, the charged geometries can have a pair of horizons in the region $r\geq \ell$. A sufficient condition to ensure the absence of such Cauchy horizon is given by $ a(\ell)<0$. Thus, as long as the following bound between the mass and the charge is satisfied,
\begin{eqnarray}
\mu > \ell^{n-1} \left( \frac{2}{3} + \frac{q^2}{\ell^{2(n-1)}} \right)\,, \label{BBnomassinflation}
\end{eqnarray}
\eqref{aBB} describes a charged regular black hole which does not suffer from the mass inflation instability. Moreover, when the charge becomes too large w.r.t. the mass, the geometry becomes horizonless, so that the mass inflation regime of these RBHs is in some sense bounded. Therefore, if mass-inflation does not drive the geometry outside of spherical symmetry, it might end-up producing a single-horizon RBH or an extremal black hole of radius $r_e$, for 
\begin{eqnarray}
\begin{split}
\mu &= \frac{1}{3} \left( 4 + \left( 2 + \frac{(2s+3)\ell^2}{(n-1)r_e^2} \right) \left( 1 - \frac{\ell^2}{r_e^2}\right)^{-\frac{1}{2}+s+n} \right) r_e^{n-1}\,,\\
q^2 &= \frac{1}{3} \left( 2 +\left( 1 + \frac{(n+2s+2) \ell^2}{(n-1)r_e^2} \right) \left( 1 - \frac{\ell^2}{r_e^2}\right)^{-\frac{1}{2}+s+n} \right) r_e^{2(n-1)}\,.
\end{split}
\end{eqnarray}

As the curvature is finite at $r=\ell$, the geometry can in principle be extended beyond this point. Furthermore, it can be checked that the differential curvature invariants $\Box^p R$ constructed using \eqref{Boxf} and \eqref{Rsf} can be made regular as well at $r=\ell$ provided that the integer $s$ be sufficiently large. For instance, $\Box R$ and $\Box^2 R$ are finite at $r=\ell$ in four dimensions $n=2$ for respectively $s\geq 2$ and $s\geq 4$, in five dimensions for respectively $s\geq 1$ and $s\geq 3$ and in higher dimensions for respectively $s\geq 0$ and $s\geq 2$. Thus, we can infer that $\Box^{p} R$ is finite for $s\geq 2(p+1)-n$.
In order to extend the geometry beyond $r=\ell$, we consider the following coordinate transformation which duplicates the $r\geq \ell$ region,
\begin{eqnarray}
r= \sqrt{\rho^2 + \ell^2}\,,
\end{eqnarray}
with $-\infty<\rho<\infty$ so that the interval becomes
\begin{eqnarray}
ds^2= - f(\rho) dt^2 + \frac{\rho^2 d\rho^2}{\left(\rho^2 + \ell^2 \right) f(\rho)} + \left( \rho^2 + \ell^2 \right) d\Omega^2_{n,1} \,, \label{x2BB}
\end{eqnarray}
where
\begin{eqnarray}
f(\rho) = 1 +\frac{3 \left(q^2 - \mu \left(\rho^2+\ell^2 \right)^{\frac{n-1}{2}} \right)\left(\rho^2+\ell^2 \right)^{\frac{3}{2}+s} }{2 \left(\rho^2+\ell^2 \right)^{\frac{1}{2}+s+n} +|\rho| \rho^{2(s+n)}} \,.
\end{eqnarray}
In particular, this type of geometry automatically makes the electric field \eqref{ElectricField} regular 
\begin{eqnarray}
\Phi = \frac{\Gamma\left(\frac{n+1}{2} \right) Q}{2 \pi^{n+1}\left( \rho^2 + \ell^2\right)^{\frac{n}{2}}}\,,  \label{ElecPotRegBB}
\end{eqnarray}
thus resolving the Coulomb singularity without the need for the non-minimal couplings studied previously in \ref{Sec.NMC}, meaning with $v=0$.

However, it can be seen from \eqref{x2BB} that the metric is degenerate at $r=0$, although the curvature invariants are regular. This is very similar to the so-called defect-wormholes \cite{Klinkhamer:2022rsj,Feng:2023vqp,Baines:2023xje}, which are quasi-regular spacetimes \cite{Ellis:1977pj} possessing a defect at the origin, making the continuation of geodesics from $r>0$ to $r<0$ at least non-unique. Indeed, in four dimensions, tidal forces along time-like geodesics remain finite at $r=\ell$ or $\rho=0$, as it can be checked deriving the components of the Riemann tensor in a parallelly propagated orthonormal frame along such curve. However, divergences might still occur along accelerated observer trajectories, such as charged particles, but this goes beyond the scope of this paper. Overall, although this quasi-regular singularity is worst than what happens for the aforementioned black bounce geometries, this still constitutes an improvement compared to the strong curvature singularities of Reissner–Nordstr\"{o}m black holes and (A)dS-core black holes suffering from mass inflation.

\subsection{Extremal inner horizon black holes with dS-cores}\label{secinnerextremal}

Let us conclude by an approach able to tame the mass inflation instability within the theory \eqref{fullaction}. The idea is to consider the series of corrections defined by \eqref{UVT1} and \eqref{UVT2} which admits an extremal black hole vacuum \eqref{SharedVac} and (A)dS-core black holes for any truncation order $m \geq 2$. Recall that this implies that Nernst's third law of thermodynamics is satisfied in this theory. As the highest curvature order $m$ in the action increases, the solution becomes smoother and the inner horizon approaches extremality for any mass, so that its surface gravity tends to vanish in the limit $m\to \infty$. We work in four dimensions and with vanishing cosmological constant.

Consider the spherical charged regular black holes whose vacuum is an extremal deformation of M$_4$, 
\begin{eqnarray}
a_m(r) = \left( \left(r^2 -  \ell^2 \right)^2  -  \left( \frac{\mu}{r} - \frac{q^2}{r^2} \right) \Pi^{\,-1}_m(r)   \right) \left( r^4 +  \ell^4  + \frac{1}{2} \beta r^2 \ell^2\right)^{-1}  \,,
\end{eqnarray}
where
\begin{eqnarray}
\Pi_m(r)=\frac{1}{\ell^4}\, \sum_{p=2}^m \sigma_p \left( \frac{ \ell^2}{r^2}\right)^p \,,
\end{eqnarray}
with $\sigma_2=1$, which are solutions of the theory \eqref{fullaction} for the couplings \eqref{UVT1}, \eqref{UVT2} and \eqref{ExtremalFT} with $\Lambda=0$. For $\beta=4$ and $\sigma_p=p-1$, it reduces the eq(214) of \cite{Colleaux:2019ckh}. However, recall from \ref{secexvac} that requiring the theory to admit a smooth hyperbolic vacuum implies $-4 <\beta < 4$. In order to have at most two horizons in the neutral case for any $m\geq2$, what is the minimal requirement for resolving curvature singularity via (A)dS-cores, we consider the simple theory \cite{Colleaux:2019ckh}
\begin{eqnarray}
\sigma_p = p-1\,, \label{pmoins1}
\end{eqnarray}
 for which the solutions read
\begin{eqnarray}
a_m(r) = \frac{\left(r^2 -  \ell^2 \right)^2}{ r^4 +  \ell^4  + \frac{1}{2} \beta r^2 \ell^2} \left( 1 - \left( \frac{\mu}{r} - \frac{q^2}{r^2} \right) \frac{1}{1 + \left( \frac{\ell}{r}\right)^{2m} \left( m-1 -m \left( \frac{r}{\ell} \right)^2  \right)} \right)\,, \label{ExInnHorm}
\end{eqnarray}
and describe charged dS$_4$-core\footnote{It is possible to have an AdS$_4$-core only for $m=2$ when the charge satisfies $q^2 > \ell^2 (4+\beta)/2$.} black holes behaving asymptotically as Reissner–Nordstr\"{o}m spacetime with a quantum-like correction to Newton's potential and admitting the above property. It can be further checked that when $q\neq 0$, there are most $4$ horizons for $m>3$. As we are interested in large order of curvature corrections, meaning in particular $m>2$, 
\begin{eqnarray}
\begin{split}
a_m\left(r\to 0\right) &=  1 -  \frac{\eta r^2}{\ell^2} +\dots\,,\\
a_m\left(r\to \infty\right) &=  1 - \frac{\mu}{r} + \frac{q^2 - \eta \ell^2}{r^2} +  \frac{\eta \mu \ell^2}{r^3} + \dots\,,
\end{split}
\end{eqnarray}
where
\begin{eqnarray}
\eta =  \frac{4 + \beta}{2}\,.
\end{eqnarray}
As explained in \ref{secsmoothness}, the geometry becomes smoother as $m$ increases, i.e. as more corrections are taken into account in the action. This is quite different from the parameter $s$ appearing in the theory \eqref{BBtheory}, which also controls the smoothness of the quasi-regular black hole \eqref{aBB}, as assuming a large value for $s$ is unnatural, while considering theories with high number of corrections is not\footnote{At least in the present context of a toy model EFT of a (perturbatively) non-renormalisable theory.}. 
In order to see the appearance of an extremal inner horizon, consider the non-perturbative ($m\to \infty$) solution 
\begin{eqnarray}
\label{NonperturbativeDiscont}
a_{\infty}  (r) =   \left\{
  \begin{array}{@{}ll@{}}
   \frac{\left(r^2 -  \ell^2 \right)^2}{ r^4 +  \ell^4  + \frac{1}{2} \beta r^2 \ell^2}   \text{\;\;\;, \;\;\;\;\;\;\;\;\;\;\;\;\;\;\;\;\;\;\;\;\;\;\;\;\;\;\;\;\;\;\;\;\;\;  if \;\; $r \leq \ell$,} 
 \\ 
 \\  \frac{\left(r^2 -  \ell^2 \right)^2}{ r^4 +  \ell^4  + \frac{1}{2} \beta r^2 \ell^2}   \left( 1 - \frac{\mu}{r} + \frac{q^2}{r^2}  \right)   \text{\;\;\;, \;\;\;\;\;\;\;\;\;\;\;\;  if \;\;  $r \geq \ell$\,,} 
  \end{array}\right.  
\end{eqnarray}
which is piecewise definite. This is a generic behaviour of this type of solutions given \eqref{EFTsmoothness} which implies that the black hole core is less and less affected by the mass as $m\to \infty$, while the asymptotic region always is, see the discussion in \ref{sec.smoothminf}.  Thus, we can deduce that the inner horizon in the neutral case and the two mostly inner horizons, when they exist for $q\neq 0$, all approach $r=\ell$ as the number of corrections grows. In particular, their surface gravity can be inferred near this limit from the generic behaviour of the solution around $r=\ell$, 
\begin{eqnarray}
\begin{split}
a_m\left(r\to \ell\right) &=- \frac{2 \mu_e}{m^2 \eta\ell} + \dots + \left( -\frac{8\mu_e}{3 m \eta \ell^2}  + \dots \right) \left(r-\ell\right)  + \left( \frac{2 \left( 18 \ell - 7 \mu_e \right)}{9 \eta\ell^3} + \dots\right) \left(r-\ell\right)^2 
\\
&+ \sum_{i\geq 3} \left(\frac{ c_i \mu_e m^{i-2}}{\eta \ell^{i+1}} + \dots \right) \left(r-\ell\right)^{i} \,,
\end{split} \label{am(r=l)}
\end{eqnarray}
which is subsequently organised according to the leading terms when $m$ is large and where the $c_i$ are numerical factors and
\begin{eqnarray}
\mu_e = \mu - \frac{q^2}{\ell}\,.
\end{eqnarray}
Therefore, the (two mostly) inner horizon(s) surface gravity is (are when they exist for $q\neq 0$) approximately given for large number of corrections by 
\begin{eqnarray}
\kappa_{-} = \frac{a_{m}'(r_{-})}{2} \approx -\frac{4\mu_e}{3 m \eta \ell^2} + \dots
\end{eqnarray}
and the mass inflation instability, which is controlled by the value of $\kappa_{-}$, can be made arbitrarily small providing that one considers theories with sufficiently high number of corrections $m$. Moreover, for any large but finite values of $m$, the geometry is effectively smooth, so that one can avoid the issue of the discontinuity of the non-perturbative solution (\ref{NonperturbativeDiscont}) and tame the mass inflation by working with very large but finite number of corrections. This provides a mechanism to tame this instability in the context of (A)dS-core regular black hole geometries.

Let us comment on the non-perturbative solution \eqref{NonperturbativeDiscont} and the associated sensitivity of the generic solution \eqref{ExInnHorm} on $m$ at $r=\ell$. It is clear that the limit $m \to \infty$ produces an ill-defined geometry at $r=\ell$, as seen from  \eqref{am(r=l)}. In particular the order at which the derivatives of the metric blow up ($i\geq 3$) are directly related to the fact that the vacuum is extremal, so that the geometry is well-defined up to $i=2$. Therefore, in order to fully avoid the mass inflation with this mechanism, what requires to consider non-perturbative theories $m\to\infty$, the vacuum should be multiply or infinitely degenerate, what also cures the Aretakis instability \cite{Aretakis:2012ei,Kehle:2022uvc,Agrawal:2026oka}. We leave this for future works.

 There can be another inner horizon in the charged case, in addition to the (near-)inner-extremal one, around which mass inflation should occur as well. Let us restrict for simplicity to the charged non-perturbative black hole \eqref{NonperturbativeDiscont}, although our conclusion are approximatively valid  for large $m$ for the generic solution as well \eqref{ExInnHorm}.  The horizons are located at $r=\ell$ and
\begin{eqnarray}
r_\pm = \frac{1}{2} \sqrt{\mu \pm \sqrt{\mu^2 - 4 q^2}} \,. 
\end{eqnarray}
Therefore, in order to avoid mass inflation, the metric should either be horizonless for $r>\ell$, or possess a single horizon, possibly extremal. The former case requires either
\begin{eqnarray}
\mu \leq 2 |q| \,, \label{muq}
\end{eqnarray}
or
\begin{eqnarray}
q \leq \ell \,,\;\;\;\;\;\;\; 2 q \leq \mu \leq \ell \left( 1 + \frac{q^2}{\ell^2} \right)\,, \label{muq2}
\end{eqnarray}
 while the latter implies
 \begin{eqnarray}
 \ell \left( 1 + \frac{q^2}{\ell^2} \right) \leq \mu  \,. \label{luq}
 \end{eqnarray}
 Reintroducing the mass \eqref{mu} and charge \eqref{redq}, restoring the units and assuming that $\ell$ is the Planck length,  the regime for which these charged black holes are stable is given by 
\begin{eqnarray}
M \geq  \frac{1}{2}\sqrt{\frac{\hbar c}{G}} \left( 1 + \frac{\alpha Q^2}{e^2} \right)\,. \label{boundle}
\end{eqnarray}
For stellar mass black holes $M \approx 10^{30} \text{kg}$, this bound gives $Q \lessapprox 18 C$, which is the typical charge of a lightning on Earth. Although adding more charge to the black hole makes it unstable, there is a maximal charge $|q|=\mu /2$ for which it becomes extremal, and above which the geometry is horizonless for $r>\ell$. Overall, the bounds \eqref{boundle} and \eqref{BBnomassinflation} (as well as eqs(282,315) of \cite{Colleaux:2019ckh} for black bounces) ensure that a  late time static charged black hole regime exists in some region of parameter space.

\section{Summary and discussion}

In this paper we have shown that when interpreted as a $d$-dimensional pure gravity action such as \eqref{fullaction}, the 2D Einstein-dilaton gravity \eqref{2Daction}, uniquely obtained from the four conditions enforcing a GR-like behaviour of the small scale curvature corrections \ref{Deftheory}, admits the following properties. Birkhoff theorem is satisfied \eqref{Eqfct} and the theory belongs to quasi-topological gravities \eqref{SFsol}. Its spherically symmetric solutions \eqref{DSS} are  exact for an arbitrary number coupling constants \eqref{RRBHs}. They possess quantum-like features and similarities with Lovelock-Lanczos gravity, such as its entropy \eqref{SLog}, and its dimensional regularisations. In particular, the critical order theory \eqref{critaction} in spacetimes \eqref{DSS} is proportional to the Euler characteristic of the 2D orbit space \eqref{EulerCh}. When regularised \eqref{RegCritL}, it yields a quantum-like correction to the Newtonian potential \eqref{QCorrNewt}, \eqref{QCorrNewtd} and logarithmic correction to the black hole entropy \eqref{SLog} in arbitrary even dimension, which also appears in the four dimensional regularisation of Gauss-Bonnet gravity \cite{Wei:2020poh}. Furthermore, any curvature correction above the critical order regularises the Schwarzschild singularity, assuming a single constraint between the couplings \eqref{RegCdt}, turning it into an (A)dS-core \eqref{AdScoreOddCrit}, \eqref{AdScore}. This is stronger but analogous to what happens in polynomial higher order gravities \cite{Giacchini:2018gxp, Giacchini:2018wlf}. Thus, contrary to Designer Lovelock Gravity \cite{Kunstatter:2015raa,Kunstatter:2015vxa} and its NPG \cite{Colleaux:2017ibe,Colleaux:2019ckh} and higher-dimensional QTG \cite{Bueno:2024dgm} formulations, regularity does not require an infinite tower of curvature corrections, but just a single non-analytic invariant, see e.g. the minimal 4D and 5D theories \eqref{4Dmin} and \eqref{5Dmin}. The charged generalisation of these solutions \eqref{RBHSTRNAdS} have been obtained and studied in \ref{secCharge}. The regularity of the black holes is preserved in the presence of charge \ref{RobRBHQ} and simple non-minimal couplings \eqref{Im} can regularise the Coulomb singularity without fine-tuning \eqref{regelec}, with exact solutions for arbitrary couplings as well \eqref{potelec}.

When considered in vacuum $M=Q=0$, these solutions are smooth deformations of maximally symmetric geometries which are recovered for $\ell \to 0$ or asymptotically $r\to \infty$, although Minkowski space can be obtained (as a limit) for theories satisfying \eqref{M4Vac}. This enabled us to construct theories \eqref{UVT1} \eqref{UVT2} admitting extremal black hole vacua \eqref{SharedVacExtr}, where the location of the horizon is fixed by the theory $r=\ell$. We have shown that this provides a concrete realisation of Nernst third law of thermodynamics, as the (minisuperspace Wald \ref{App}) entropy associated with these horizons is a universal constant \eqref{Svac}.
We interpreted these vacua as a way to implement a minimal resolution scale in spherically symmetric spacetimes, given that the localisation of test particles below $r=\ell$ is lost for the external observer.

We addressed the issues of exact smoothness and mass-inflation in \ref{Sec.inf}, considering infinite towers of curvature corrections $m \to \infty$. Indeed, while odd-dimensional solutions \eqref{RBHSTAdS} are smooth, even-dimensional geometries approach smoothness when the highest order of correction in the action $m \to \infty$, without any fine-tuning \eqref{EFTsmoothness}. Although we argued that this notion of smoothness is more robust, we constructed theories with exact smooth Bardeen-type black holes \eqref{Bardeen}, \eqref{smooth2}. Furthermore, (A)dS-core regular black holes necessarily possess an inner horizon, unstable against mass-inflation. We proposed two mechanisms to ensure that late time static black hole regime exists in some region of parameter space. First, we constructed a theory \eqref{BBtheory} admitting an exact quasi-regular charged black hole geometry \eqref{aBB}, which possesses a single-horizon in the uncharged case or when some bound between the mass and charge is satisfied \eqref{BBnomassinflation}. Interestingly, a similar bound \eqref{boundle} has been obtained in our second mechanism to avoid mass inflation, based on black hole solutions \eqref{ExInnHorm}  with an extremal inner horizon in the non-perturbative limit $m\to \infty$ \eqref{NonperturbativeDiscont}. These two different geometries, as well as the charged black bounces obtained from NPG actions in \cite{Colleaux:2019ckh}, indicate that for a stellar mass charged black holes, the mass inflation is avoided when the charge satisfies a bound of the order of $Q \lessapprox 10 C$.

\medskip

These properties and the wide variety of regular black holes which can be constructed from the 2D reduction \eqref{2Daction} of the theory \eqref{fullaction} shows a flexibility which would be interesting to extend systematically to dynamical spherically symmetric black holes, using the reduced field equations \eqref{2DcovEOM} with more general matter content. Indeed, this framework enable to study in details gravitational collapse, black hole evaporation, primordial black holes and remnants as dark matter candidates and mass inflation \cite{Maeda:2006pm,Maeda:2011ii,Ohashi:2011zza,Biasi:2022ktq,Bueno:2024eig,Bueno:2025zaj,Barenboim:2025ckx,Khlopov:2008qy,Dymnikova:2015yma,Calza:2024fzo, Calza:2024xdh, Calza:2025mwn}. It would be interesting to perform a comparison analysis of these phenomena for smooth black holes obtained from non-perturbative theories $m\to\infty$, as well as for (A)dS-core black holes obtained from large order theories, whose regularity is more robust to the inclusion of matter. In particular, it would be desirable to obtain the minimal number of corrections needed to cure the singularity for all form of physically relevant matter satisfying suitable energy conditions. Notice that a selection rule for effective 2D Horndeski and their NPG formulation has recently been proposed in \cite{Thaalba:2026abz},  based on their ability to support dynamical regular matter configurations. As it turns out, our 4D perturbative solutions \eqref{RBHSTAdS}, \eqref{RBHSTRNAdS}, \eqref{ExInnHorm} and \eqref{NonperturbativeDiscont} satisfy this parity rule $a(-r,-M)=a(r,M)$, while the smooth black holes \eqref{Bardeen} and \eqref{smooth2} do not. The case of the quasi-regular geometries would also be worth investigating further, as, if singularities are cured by quantum gravity and this resolution survives in some effective semi-classical regime, it might be that the former leaves some imprints on the latter, for instance in the form of quasi-regular singularities and topological defects, with possible observational consequences \cite{Krasnikov:2009xt,Stadnik:2014cea}.

As mentioned previously, the theories \eqref{pmoins1} admitting extremal vacuum and near-extremal-inner horizon for $m \gg 1$ should be possible to refine using degenerate vacuum with higher or infinite multiplicity, so that the non-perturbative limit should be well-defined and mass inflation strictly avoided. A dynamical study of the theory \eqref{pmoins1} and such extensions should enable to analyse in details the status of Aretakis and mass inflation instabilities in these models and compare with recent results on the UV sensitivity of extremal black holes  \cite{Horowitz:2023xyl,DelPorro:2025fiu}. Together with a study of evaporation, this should enable to assess whether degenerate vacua are viable candidate to describe massless small scale deformations of maximally symmetric spaces.

As explained below \eqref{gamma4d}, the non-analyticity of non-polynomial gravity actions implies that conformally flat geometries are not exact solutions of these models but only limiting ones. Thus, theories with non-maximally symmetric vacua such as extremal \eqref{SharedVacExtr} and multi/infinitely-degenerate ones might be preferred. This also means that, strictly speaking, cosmology should be studied using anisotropic geometries in these theories and that collapse should not presuppose a conformally flat interior region. The generalisation of NPG to more general backgrounds, such as those admitting the symmetries of near-horizon extremal rotating Kerr-Newman black hole \cite{Colleaux:2026qew}, seems to be a natural avenue forward, as the interesting physics of NPG adapted to spherical symmetry is already (almost entirely) captured in their associated 2D Horndeski reduction, while it remains unclear what types of rotating black holes and nutty geometries could emerge from 4D diffeomorphism invariant pure gravity. If such generalisations are possible, the complete integrability of the theory \eqref{fullaction} in spherical symmetry \eqref{RRBHs} would certainly help in obtaining exact solutions for spacetimes with less symmetries. Finally, as an approach aimed at describing some exotic gravitational EFT around strong field configurations, it would be interesting to assess whether some NPG representatives (see \ref{Dico}) exist such that perturbations around black hole spacetimes are well-defined and second order, as it happens at linear order for QTG and some NPG around maximally symmetric spaces \cite{Bueno:2025zaj}.

\begin{acknowledgments}
We would like to thank Sergio Zerbini, Stefano Chinaglia, Luciano Vanzo and Massimiliano Rinaldi for very useful discussions during my PhD, where most of these results have been obtained, and to Karim Noui and Gonzalo J. Olmo for useful comments on my thesis. We thank as well Zakaria Belkhadria, Ivan Kolář, Pavel Krtouš, David Kubizňák and Alejandro Perez for interesting recent discussions and references.
 We acknowledge financial support from the Primus grant PRIMUS/23/SCI/005 of Charles University and the Charles University Research Center grant UNCE24/SCI/016.
\end{acknowledgments}

\appendix

\section{Curvature of spherical (hyperbolic, planar) symmetric geometries}\label{appci}

The independent components of the $d$-dimensional Riemann and Weyl tensors in spacetimes \eqref{DSS} are given by 
\begin{eqnarray}
\begin{split}
R^{ai}_{\;\; bj} &= -\frac{D^a D_b r}{r} \, \delta_i^j \,,\;\;\;\;\;\;\;\;\; R^{ij}_{\;\;kl} = \frac{k- D^a r D_a r}{r^2} \, \delta^i_{[k} \delta_{l]}^j\,,\;\;\;\;\;\;\;\;\; R^{ab}_{\;\;ce} = \frac{1}{2} R\left( \gamma \right) \delta^a_{[c}\delta_{e]}^b \,,\\
W^{ab}_{\;\; ce} &= \frac{d-3}{d-1} \delta^{a}_{[c}\delta^{b}_{e]} \psi  \,,\;\;\;\;\;\;\; W^{ij}_{\;\; kl} = \frac{2}{(d-2)(d-1)} \delta^i_{[l} \delta^j_{k]} \psi \,,\;\;\;\;\;\;\; W^{ai}_{\;\; bj} = \frac{d-3}{(d-2)(d-1)} \delta^a_b \delta^i_j \psi \,,
\end{split}
\label{Riemcomp}
\end{eqnarray}
where 
\begin{eqnarray}
\psi = - \left( \frac{1}{2} R\left(\gamma\right)  + \frac{D_a D^a r}{r} +  \frac{k- D^a r D_a r}{r^2}  \right).
\end{eqnarray}
Using the identity $C^{\mu\nu}_{\;\; \alpha} = - \frac{d-2}{d-3} \nabla^\beta W^{\mu\nu}_{\;\; \alpha\beta}$ leads to the decomposition of the $d$-dimensional Cotton tensor 
\begin{eqnarray}
C^{ec}_{\;\; b} = (d-2) \left( \xi^{[c} \delta^{e]}_b \psi + \frac{1}{d-1} \delta^{[e}_b \partial^{c]} \psi \right) \, , \;\;\;\;\;\;\;\;\;\;
C^{jb}_{\;\; i} = - \delta_i^j \left( \frac{1}{d-1} \partial^b \psi + \xi^b \psi\right) \, ,
\end{eqnarray}
where $\xi_a = D_a \log r$. Equation \eqref{uDecomp} follows from the squares of this tensor, whose decompositions read
\begin{eqnarray}
\begin{split} 
\label{Cotton2D}
C^{\mu\nu}_{\;\; \alpha} C_{\mu\nu}^{\;\; \alpha} &= 2 (d-2) (d-1) \Omega_a \Omega^a \, , \\
C^{\mu\,\, \beta}_{\; \alpha} C^{\nu \alpha}_{\;\;\beta} &= (d-2) \delta_a^\mu \delta_b^\nu \Omega^a \Omega^b +\Big( \delta_i^\mu \delta_j^\nu \sigma^{ij} + (d-2)^2  \delta_a^\mu \delta_b^\nu \gamma^{ab} \Big) \Omega^c \Omega_c \, , \\
C^{\mu\alpha}_{\;\; \beta} C^{\nu\beta}_{\;\;\alpha} &= (d-2)(d-1) \delta_a^\mu \delta_b^\nu \Omega^a \Omega^b \, .
\end{split}
\end{eqnarray}
where $\Omega_a = \frac{\partial_a \psi}{d-1} + \xi_a \psi$.

\medskip

 Using the scalar field as radial coordinate, the solutions of the theory \eqref{fullaction} can be written as
\begin{eqnarray}
d\Sigma^2 = - a(r) dt^2 + \frac{dr^2}{a(r)}\,, 
\end{eqnarray}
so that 
\begin{eqnarray}
R^{ai}_{\;\; bj} = - \frac{a'}{2r} \, \delta^a_b  \, \delta_i^j \,,\;\;\;\;\;\;\;\;\; R^{ij}_{\;\;kl} = \frac{k- a}{r^2} \, \delta^i_{[k} \delta_{l]}^j\,,\;\;\;\;\;\;\;\;\; R^{ab}_{\;\;ce} = - \frac{1}{2} a'' \, \delta^a_{[c}\delta_{e]}^b \,.
\label{RiemcompRed}
\end{eqnarray}
As we consider geometries which are locally asymptotically (A)dS or Minkowski, a sufficient condition to have regular polynomial invariants in the Riemann tensor, is that $a$, $(k- a)/r^2$, $a'/r$ and $a''$ be bounded for finite $r$. As far as differential invariants are concerned, we use the following probe \cite{Giacchini:2021pmr}. Given a scalar invariant $f(r)$, we consider the $d$-dimensional Laplace-Beltrami operator which is given for the previous geometries by
\begin{eqnarray}
\Box f =\left( D^2  + n  \gamma^{ab}\, \frac{D_a r \, \partial_b}{r}   \right) f = \left( \left(  a'+ \frac{ n a}{r}  \right) \frac{d}{dr} + a \frac{d^2}{dr^2} \right) f  \,, \label{Boxf}
\end{eqnarray}
where $D^2$ is the two-dimensional one. Applying this operator multiple times on the $d$-dimensional Ricci scalar,
\begin{eqnarray}
R = - a'' - \frac{2n a'}{r} + n(n-1) \frac{k-a}{r^2}  \,, \label{Rsf}
\end{eqnarray}
or any other curvature invariants, enables to probe the smoothness of the geometry in a covariant way.

\section{Minisuperspace Wald entropy}\label{App}

For a given diffeomorphism invariant gravitational theory with Lagrangian density $L$, the Wald-Iyer entropy \cite{Wald:1993nt,Iyer:1994ys}, is given by  the following integral over a cross-section $\Omega$ of the event horizon 
\begin{eqnarray}
 S_{W}=-2\pi \oint_{\Omega} \frac{\delta L}{\delta R_{\mu\nu\alpha\beta}} \varepsilon_{\mu\nu} \varepsilon_{\alpha\beta} d\Omega_r\,, \label{WaldEntropy}
\end{eqnarray}
where $\varepsilon_{\mu\nu}$ is the binormal to $\Omega$, satisfying $\varepsilon_{\mu\nu} \varepsilon^{\mu\nu} =-2$ and $d\Omega_r$ is its the volume form. It can be applied for asymptotically flat spacetime admitting a bifurcate Killing horizon. Our action \eqref{fullaction} being diffeomorphism invariant, one could in principle derives its entropy using the previous formula. However, we have proposed in \cite{Colleaux:2019ckh} a simpler minisuperspace formula for the Wald entropy which can be applied for spacetimes \eqref{DSS}, and checked in various situations, such as Lovelock-Lanczos gravity, quasi-topological gravity, generalised quasi-topological gravity \cite{Hennigar:2017ego}, and some genuine non-polynomial gravity actions \cite{Deser:2007za, Bellini:2010ar} that both notions of entropy agree (see Appendix C. of \cite{Colleaux:2019ckh}). It is simply given by
\begin{eqnarray}
 S =-2\pi \oint_\Omega \frac{\delta L\big|}{\delta R_{abcd}}  \varepsilon_{ab} \varepsilon_{cd} \, d\Omega_r  \,,\label{Smini}
\end{eqnarray}
where the vertical bar denote the evaluation of the Lagrangian density on the ansatz \eqref{DSS}, owing to the fact that the binormal is by definition transverse to the horizon.

Remark that its agreement with \eqref{WaldEntropy} is not surprising from the point of view of the principle of symmetric criticality \cite{Palais:1979rca,Torre:2010xa,Fels:2001rv,Frausto:2024egp}, assessing for which symmetry the variation and symmetry reduction commute. In particular it is valid for the symmetries of \eqref{DSS}. Given that non-polynomial gravity actions are usually designed to yield second order field equations in \eqref{DSS}, the previous minisuperspace entropy is well-adapted to study their thermodynamics. In particular, assuming a reduced Lagrangian of the form 
\begin{eqnarray}
L\big| = \mathscr{L}\left( R\left(\gamma\right), \gamma_{ab}, r, D_a r , \dots, D_{(a_1} \dots D_{a_i)} r  \right)
\end{eqnarray}
and using that the components of the $d$-dimensional Riemann tensor in the 2D base space are given by \eqref{Riemcomp}, we obtain
\begin{eqnarray}
 S= 4 \pi \oint_\Omega \frac{\partial L\big|}{\partial R\left(\gamma\right)}  \, d\Omega_r  \,.
\end{eqnarray}
Considering a static black hole metric \eqref{DSS} in a gauge in which the scalar field $r$ is the radial coordinate, with $d\Sigma^{\,2} = -a(r) dt^2 + a(r)^{-1} dr^2$, $\Omega$ is given by the surface of constant time $t$ at the horizon $r=r_H$, obtained by solving $a\left(r_H \right)=0$. Thus
\begin{eqnarray}
 S= 4  \pi  \mathcal{A}_k \left. \frac{\partial L\big|}{\partial R\left(\gamma\right)}\right|_{r=r_H}    \,, \label{Sms}
\end{eqnarray}
where $\mathcal{A}_k = r_H^n V_{n,k}$ is the area of the black hole horizon.

Notice that there are some ambiguities in  \eqref{WaldEntropy} and \eqref{Sms}. For instance, the 2D Ricci scalar can be locally written as a divergence in 2D \eqref{RicciD}. Thus, any action linear in this term can be equivalently written without it, modulo boundary terms, without introducing additional structure than the scalar field $r$ \cite{Colleaux:2019ckh}. This is of course expected, as boundary terms modify the symplectic structure, from which the formula \eqref{WaldEntropy} is derived. Ambiguities of \eqref{Smini} for non-stationary horizons should also be expected, as they are already present  in \eqref{WaldEntropy}, see \cite{Jacobson:1993vj}. However, given that the theory preserves its analyticity and second order field equations for any dynamical metrics \eqref{DSS} and resembles Lovelock gravity for these geometries, we expect that the dynamical entropy defined through a unified first law, as in  \cite{Nozawa:2007vq,Maeda:2011ii}, should agree with our result \eqref{SLog}. Finally, although extremal horizons are not bifurcate Killing, one may derive the Wald entropy from the non-degenerate black hole and continuously take the extremal limit, as it is often done.

\section{Charged regular black hole solutions}\label{secCharge}

In order to assess the robustness of the corrections \eqref{fullaction} to resolve curvature singularities, we introduce an electromagnetic field and show that charged black holes also admit an (A)dS-core.  This is a crucial criterion to select theories with regular black holes, see \cite{Colleaux:2019ckh} and more recently \cite{Carballo-Rubio:2026mvj}.

Furthermore, we also consider a non-minimal coupling in the form of terms $\mathcal{R}^{p} F^{\mu\nu}F_{\mu\nu}$, where $\mathcal{R}$ is defined by Eq(\ref{RS}). Recall that in spherically symmetric spacetimes, this scalar reduces to the Ricci scalar of the horizon manifold, and it is the same that we have used to produce higher order corrections to the Einstein-Hilbert action. As we will see, the presence of this simple term regularizes the Coulomb singularity of the electric field in spherically symmetric spacetimes, provided that the order of correction is at least critical, $p \geq \frac{n}{2}$, as it also happens for curvature singularities.

\subsection{Maxwell field non-minimally coupled to the curvature $\mathcal{R}$}

Let us introduce the U(1) gauge field $A_\mu$ with its field strength $F_{\mu\nu}= \partial_\mu A_\nu - \partial_\nu A_\mu$, and consider the curvature-corrected non-minimally coupled matter action,
\begin{eqnarray}
I_m = -\frac{1}{4} \int_\mathcal{M} d^{\, n+2}x  \, \sqrt{-g} \, F^{\mu\nu}F_{\mu\nu} \sum _{p=0}^v \xi_p \left( \ell^{2} \mathcal{R}\right)^p\,, \label{Im}
\end{eqnarray}
where $\xi_0=1$, so that Maxwell theory is recovered for $\ell=0$. The field equation for the gauge field reads 
\begin{eqnarray}
\nabla_\mu \left( F^{\mu\nu}\, \sum _{p=0}^v \xi_p \left( \ell^{2} \mathcal{R}\right)^p \right) =0  \,,
\end{eqnarray}
while the components of the stress-energy tensor \eqref{FieldEq}, assuming a gauge potential and Faraday tensor compatible with the symmetries of \eqref{DSS},
\begin{eqnarray}
A_\mu dx^\mu = A_a\left(x^c\right) dx^a \,,\;\;\;\;\;\;\;\;\;   F_{ab}= \Phi \,\varepsilon_{ab} \,,\;\;\;\;\;\;\;\;\;  \Phi = - \varepsilon^{ab}\, \partial_a A_b \,,
\end{eqnarray}
are given by
\begin{eqnarray}
T^a{}_b = - \frac{1}{2} \Phi^2 \delta^a{}_b \,,\;\;\;\; \;\;\;\;  T^i{}_j =  \frac{1}{2} \Phi^2 \delta^i{}_j \,.
\end{eqnarray}

\subsection{Birkhoff theorem and charged black hole solutions}

Assuming a gauge \eqref{Weylgauge} for the 2D metric on $\Sigma$, using the scalar field $r$ as radial coordinate and considering the electric potential
\begin{eqnarray}
A= \varphi(t,r) dt \,, \;\;\;\;\;\; \Phi = - b^{-1} \varphi' \,, \label{Ansatzgf}
\end{eqnarray}
the gauge field equations reduce to 
\begin{eqnarray}
\dot{\Phi} \sum_{p=0}^v \Delta_\xi  =0\,,\;\;\;\;\;\;\;\; \left( r^n\, \Phi \, \sum\limits_{p=0}^v \Delta_\xi  \right)'=0  \,.
\end{eqnarray}
where $\Delta_\xi$ is defined by \eqref{Delta}. The first equation shows the time-independence of the electromagnetic field, while the second one yields 
\begin{eqnarray}
\Phi = \frac{Q}{r^n V_{n,k}} \left( 1 + \sum\limits_{p=1}^v \Delta_\xi \right)^{-1}\,, \label{ElectricField}
\end{eqnarray}
where $Q$ is the electric charge. The first two equations of \eqref{Eqfct} are unmodified and show once again that Birkhoff theorem and the quasi-topological property are satisfied, meaning $b=1$ and $a=a(r)$. The remaining equation \eqref{FieldEq} reads
\begin{eqnarray}
\frac{k}{r^2}\sum_{p=-1}^m \Delta_\alpha - \frac{1}{r^n} \left(   r^{n-1} \, a \sum_{p=-1}^m \Delta_\beta  \right)'= 8 \pi G  \,\Phi^2  \, \sum_{p=0}^v \Delta_\xi  \,.
\end{eqnarray}
Using \eqref{ElectricField} and \eqref{Ansatzgf}, this equation can be integrated 
\begin{eqnarray}
\int \frac{k}{r}\sum_{p=-1}^m \Delta_\alpha -     \, a \sum_{p=-1}^m \Delta_\beta  =\frac{ n \mu }{r^{n-1}}  -  \frac{ 8 \pi G  Q  }{V_{n,k}}\frac{\varphi}{r^{n-1}}  \,,
\end{eqnarray}
so that charged black hole solutions can be found from neutral ones \eqref{RRBHs} replacing 
\begin{eqnarray}
\mu \longrightarrow  \mu - \frac{8 \pi G  Q }{n V_{n,k}}\, \varphi\,. \label{NeutralToCharged}
\end{eqnarray}
For Maxwell theory ($v=0$), it reduces to 
 \begin{eqnarray}
\mu \longrightarrow  \mu -\frac{q^2}{r^{n-1}}\,, \label{NeutralToChargedMax}
\end{eqnarray}
which is the same replacement as in Lovelock gravity \cite{Maeda:2011ii}, where we define the reduced charge squared as
\begin{eqnarray}
q^2 = \frac{8 \pi G Q^2}{ n(n-1) V_{n,k}^2} \,. \label{redq}
\end{eqnarray}

\subsection{Regularisation of Coulomb's potential from non-minimal couplings}\label{Sec.NMC}

Making use of \eqref{ElectricField}, the electromagnetic invariant $\Phi$ behaves near the origin as 
\begin{eqnarray}
\Phi \left( r\to 0 \right) = \frac{Q}{V_{n,k}} \frac{r^{2v-n}}{\xi_v \, k^v \ell^{2v}} + \dots\,. \label{regelec}
\end{eqnarray}
Therefore, the electric field can be made regular at the origin provided that the number of the non-minimally coupled  curvature corrections is at least critical,
\begin{eqnarray}
v \geq \frac{n}{2}\,.
\end{eqnarray} 
Interestingly, an exact solution for the electric potential is possible to obtain for an arbitrary number of coupling constants. Indeed, using Vieta's formula by writing $\xi$ in terms of the elementary symmetric polynomials $e$ of an auxiliary set of couplings $\omega_p$, for $p \in [1,v]$,
\begin{eqnarray}
\xi_p= (-1)^p e_{v-p}\left(-\omega\right) e_v\left(-\omega\right)^{-1}\,,
\end{eqnarray}
factorises the denominator of \eqref{ElectricField}, with roots $-\omega_p$. The general solutions can then be written as
\begin{eqnarray}
\varphi(r)= \frac{Q}{(n-1)r^{n-1} V_{n,k}}\, H_v(r)\,, \label{potelec}
\end{eqnarray}
where
\begin{eqnarray}
H_v(r) =    \sum_{p=1}^v \, \frac{-\left(-1\right)^v}{\omega_p \prod\limits_{i\neq p} \left( \omega_p - \omega_i \right) }\, {}_2F_1 \left(1 , \frac{n-1}{2} , \frac{n+1}{2} , -\frac{k}{\omega_p}\left(\frac{\ell}{r}\right)^2 \right) \prod\limits_{i=1}^v w_i\,,
\end{eqnarray}
where $i \in [1,v]$, in terms of Gauss' hypergeometric function ${}_2F_1$, which in four dimensions reduces to ${}_2F_1 \left(1 , \frac{1}{2} , \frac{3}{2} , -x \right)= \arctan \left(\sqrt{x}\right)/\sqrt{x}$. We may choose the convention  $H_0 =1$, so that $v=0$ reduces to the $d$-dimensional Coulomb potential.

Therefore, the black hole solutions of this model are also exact for arbitrary couplings and obtained from \eqref{RRBHs} using the replacement \eqref{NeutralToCharged}.

\subsection{Robust regular black holes against Maxwell field}\label{RobRBHQ}

Let us now restrict to Maxwell theory, $v=0$, which is the matter action with the most divergent electric field among \eqref{Im}, as seen from \eqref{regelec}. This enables to assess whether the resolution of curvature singularities from the corrections \eqref{fullaction} is robust when a physical matter field is included into the system. Thus, it is an important criterion to select theories admitting regular black holes.

\subsubsection{Regular charged black holes}

The charged generalisation of the regular black holes \eqref{RBHSTAdS} are obtained from \eqref{NeutralToCharged} and read
\begin{eqnarray}
a(r)= k - \frac{\frac{2 \Lambda}{n\left(n+1\right)} \left(r^2 - \frac{k \ell^2}{n^2} \sum\limits_{p=1}^{\lfloor \frac{n-2}{2} \rfloor} \Delta_\zeta \right) +\frac{\mu}{r^{n-1}} - \frac{q^2}{r^{2(n-1)}}+  \frac{k}{n}\left(\bar{\Delta}-  \Delta \log\left(\frac{r}{r_0}\right) + \sum\limits_{p=\lceil \frac{n}{2} \rceil}^{m-1}  \Delta_\gamma \right)}{1+\frac{1}{n}\sum\limits_{p=1}^m \Delta_\beta} \label{RBHSTRNAdS}
\end{eqnarray}
where the deltas are given by \eqref{barDelta}, \eqref{Delta}. Recall that in the neutral case, the condition for regularity is simply
\begin{eqnarray}
m > \frac{n}{2}\,,
\end{eqnarray}
and in odd dimensions, considering the lowest maximal order of correction $m = \left\lceil \frac{n}{2} \right\rceil$, requires to further set $\Delta=0$. The presence of the electric field associated with a point-like charge naturally  requires stronger conditions, as in the central region, charges usually generate stronger gravitational fields than masses.

 However, it is always possible to obtain regular black holes from finite order corrections. In particular, the lowest order for which the singularity can be cured is now given by
\begin{eqnarray}
m = n \,,
\end{eqnarray}
so that the black holes possess a charged (A)dS-core given by 
\begin{eqnarray}
a(r\to 0) = k - \left( \frac{\gamma_{m-1} }{ \beta_m} - \frac{n q^2}{\beta_n k^n \ell^{2(n-1)}} \right) \frac{r^2}{\ell^2} + \dots\,.\label{QAdScore1}
\end{eqnarray}
Any correction of greater order,
\begin{eqnarray}
m > n \,, 
\end{eqnarray}
produces the same vacuum-(A)dS-core as in the neutral case \eqref{AdScore}, i.e.
\begin{eqnarray}
a(r\to 0) = k - \left( \frac{\gamma_{m-1} }{ \beta_m}  \right) \frac{r^2}{\ell^2} + \dots\,.\label{QAdScore2}
\end{eqnarray}
Therefore, the resolution of curvature singularities using action \eqref{fullaction} is robust to the inclusion of matter, at least in the form of an electromagnetic field. Finally, remark that considering the curvature corrections \eqref{Im} with their associated electric potential \eqref{potelec} would weaken the previous regularity condition, yielding $\frac{n}{2}<m \leq n$ for the lowest order of correction able to cure the black hole singularity.

\subsubsection{Minimal four \& five dimensional theories}

Let us briefly see the implications of these results in four and five dimensions. To summarise from the beginning : assuming a NPG theory whose 2D reduction is the minimal extension of GR to higher orders, in the precise sense of \eqref{LCdt} and \eqref{GCdt}, yields \eqref{2Daction}, whose black hole solutions are regular provided $\gamma_m=0$ and $m>n/2$. In order for their charged generalisations to be regular as well, we need $m\geq n$, which in four and five dimensions is identical to the vacuum case. Therefore, respectively cubic and quartic curvature corrections are sufficient to cure the singularity of charged static black holes in these dimensions. However, requiring the curvature invariants to be bounded by a universal constant at the center $r=0$, i.e. requiring a vacuum-(A)dS-core for the charged black holes \eqref{QAdScore2}, implies $m>n$. Thus, in four and five dimensions, the minimal corrections are respectively quartic and quintic in curvature. Demanding for the solutions to be regular for both spherical and hyperbolic topologies implies that the correction should also be quintic in curvature in four dimensions.

 In particular, the charged generalisation of \eqref{5Dminsol}, obtained using \eqref{NeutralToChargedMax}, satisfies all these requirements,  so the five-dimensional theory \eqref{5Dmin} is still the minimal one with these properties. There are various four dimensional quintic theories admitting these properties. An example of these is given by 
\begin{eqnarray}
I_{4D} = \frac{1}{16 \pi G}  \int d^4 x \sqrt{-g} \left(-2 \Lambda+  R + \beta_1  \left( \ell^{2}  \mathcal{R} \right)  \mathscr{S} + \alpha_3\, \ell^6 \mathcal{R}^4 - \beta_{4} \left( \ell^{2}  \mathcal{R} \right)^{4} \left( 7 \mathcal{R}+\frac{\mathcal{S}_{4}}{6} \right) \right)
\end{eqnarray}
and admits the following charged black hole solutions with quantum-like correction to the Newtonian potential and logarithmic correction to the entropy,
\begin{eqnarray}
a(r)= k - \frac{r^4 \Big( \frac{1}{3} r^4 \Lambda - q^2 + \mu r + \frac{1}{2}\beta_1 k^2 \ell^2 \Big)+\frac{1}{10} \alpha_3 k^4 \ell^6}{r^8 +\frac{1}{2} k \left( \beta_1 r^6 + \beta_4 k^3 \ell^6 \right)\ell^2 }\,,
\end{eqnarray}
which are regular for $k=\pm 1$ assuming $\beta_4 > \frac{27}{2048} \,\beta_1^4$.

 \bibliography{RRBGNPGBiblio.bib}

\end{document}